\documentclass[]{aa}  

\usepackage{graphicx}
\usepackage{multirow}
\usepackage{txfonts}
\usepackage[]{hyperref}
\begin{document}

\title{{\tt p-winds}: An open-source Python code to model planetary outflows and upper atmospheres\thanks{The source code can be freely obtained in \url{https://github.com/ladsantos/p-winds}. Documentation, installation instructions, and tutorials are available in \url{https://p-winds.readthedocs.io/}. Contributions to the project are welcome.}}
\titlerunning{An open-source code to model planetary outflows and upper atmospheres}
\authorrunning{L. A. dos Santos et al.}

\author{Leonardo A. Dos Santos\inst{1,2}
\and
Aline A. Vidotto\inst{3,4}
\and
Shreyas Vissapragada\inst{5}
\and
Munazza K. Alam\inst{6,7}
\and
Romain Allart\inst{8}
\and
Vincent Bourrier\inst{2}
\and
James Kirk\inst{6}
\and
Julia V. Seidel\inst{2,9}
\and
David Ehrenreich\inst{2}
}

\institute{Space Telescope Science Institute, 3700 San Martin Drive, Baltimore, MD 21218, USA\\
\email{ldsantos@stsci.edu}
\and
Observatoire astronomique de l’Université de Genève, Chemin Pegasi 51, 1290 Versoix, Switzerland
\and
Leiden Observatory, Leiden University, Postbus 9513, 2300 RA Leiden, The Netherlands
\and
School of Physics, Trinity College Dublin, the University of Dublin, College Green, Dublin-2, Ireland
\and
Division of Geological and Planetary Sciences, California Institute of Technology, 1200 East California Blvd, Pasadena, CA 91125, USA
\and
Center for Astrophysics | Harvard \& Smithsonian, 60 Garden Street, Cambridge, MA 02138, USA
\and
Earth and Planets Laboratory, The Carnegie Institution for Science, 5241 Broad Branch Road, Washington, DC 20015, USA
\and
Department of Physics, and Institute for Research on Exoplanets, Universit\'e de Montr\'eal, Montr\'eal, H3T 1J4, Canada
\and
European Southern Observatory, Alonso de C\'ordova 3107, Santiago, Casilla 19001, Chile
}

\date{Received 17 August 2021; accepted 30 November 2021}
 
\abstract{Atmospheric escape is considered to be one of the main channels for evolution in sub-Jovian planets, particularly in their early lives. While there are several hypotheses proposed to explain escape in exoplanets, testing them with atmospheric observations remains a challenge. In this context, high-resolution transmission spectroscopy of transiting exoplanets for the metastable helium triplet (He~2$^3$S) at $1\,083$~nm has emerged as a reliable technique for observing and measuring escape. To aid in the prediction and interpretation of metastable He transmission spectroscopy observations, we developed the code {\tt p-winds}. This is an open-source, fully documented, scalable Python implementation of the one-dimensional, purely H+He Parker wind model for upper atmospheres coupled with ionization balance, ray-tracing, and radiative transfer routines. We demonstrate an atmospheric retrieval by fitting {\tt p-winds} models to the observed metastable He transmission spectrum of the warm Neptune HAT-P-11~b and take the variation in the in-transit absorption caused by transit geometry into account. For this planet, our best fit yields a total atmospheric escape rate of approximately $2.5 \times 10^{10}$~g~s$^{-1}$ and an outflow temperature of $7200$~K. The range of retrieved mass loss rates increases significantly when we let the H atom fraction be a free parameter, but its posterior distribution remains unconstrained by He observations alone. The stellar host limb darkening does not have a significant impact on the retrieved escape rate or outflow temperature for HAT-P-11~b. Based on the non-detection of escaping He for GJ~436~b, we are able to rule out total escape rates higher than $3.4 \times 10^{10}$~g~s$^{-1}$ at 99.7\% (3$\sigma$) confidence.}

\keywords{Methods: numerical -- Planets and satellites: atmospheres}

\maketitle

\section{Introduction}

The evolution of short-period exoplanets is thought to be dictated by atmospheric escape. This conclusion is supported by two different approaches: i) the detection of planetary outflows and large escape rates in hot exoplanets \citep[e.g.,][]{2003Natur.422..143V, 2015Natur.522..459E, 2018A&A...620A.147B} and ii) the observation of demographic features possibly carved by atmospheric escape in the population of Neptunes and super-Earths \citep{2013ApJ...763...12B, 2013ApJ...775..105O, 2016A&A...589A..75M, 2017AJ....154..109F, 2018AJ....156..264F, 2020ApJS..247...28H}. These discoveries have led the community to attempt to combine the theoretical descriptions of escape based on demographic features to predict observable atmospheric signatures in transiting exoplanets \citep[e.g.,][]{2007A&A...461.1185L, 2016A&A...586A..75S, 2019MNRAS.484L..49K, 2020MNRAS.498L..53C}. This experiment has been challenging, mainly because of limitations in our instruments and our theories \citep[e.g.,][]{2017MNRAS.466.1868C, 2020AJ....160..258K, 2020MNRAS.498L.119G, 2021JGRE..12606639B}. 

There are four known spectroscopic windows for observing atmospheric escape: the Lyman-$\alpha$ line at 121.57~nm \citep{2003Natur.422..143V}, metallic chromospheric lines and continuum in the ultraviolet \citep{2010ApJ...714L.222F, 2019AJ....158...91S}, the metastable helium triplet at 1\,083~nm \citep{2000ApJ...537..916S, 2018ApJ...855L..11O}, and the Balmer series of H lines in the blue optical \citep{2012ApJ...751...86J, 2020A&A...638A..87W}. Each one of them has its own set of challenges. While UV observations have classically been used to this end with a variable degree of success \citep[e.g.,][]{2010A&A...514A..72L, 2010ApJ...714L.222F, 2013A&A...560A..54V, 2019AJ....158...50W, 2020A&A...634L...4D, 2021A&A...649A..40D, 2021A&A...650A..73B, 2021ApJ...907L..36G}, they are particularly complicated because only the {\it Hubble Space Telescope} ({\it HST}) can access this wavelength range at high spectral resolution; in addition, cool stars usually do not have UV continuum, limiting transmission spectroscopy only to chromospheric or transition-region emission lines whose count rates are very low \citep{2017A&A...599L...3B, 2019A&A...629A..47D}. 

One of these techniques, He transmission spectroscopy, has been shown to be reliable and attainable using ground- and space-based instruments \citep{2018Natur.557...68S, 2018Sci...362.1384A}. This spectral channel is not photon-starved and is devoid of interstellar medium absorption \citep{2009ApJ...703.2131I}, the main limitations of Lyman-$\alpha$ spectroscopy. The disadvantage is that the formation of metastable He in the upper atmospheres of exoplanets depends on a specific level of irradiation arriving at the planet \citep[e.g.,][]{2018Sci...362.1388N, 2019ApJ...881..133O, 2020A&A...640A..29D}. Nevertheless, He spectroscopy has the potential to become the main technique of atmospheric escape observations \citep{2019A&A...623A..58A, 2019A&A...629A.110A, 2020AJ....159..115K, 2020AJ....159..278V, 2021A&A...647A.129L, 2021ApJ...909L..10P}.

Upper atmospheres extend to several planetary radii and can dwarf the size of planet-hosting stars depending on the properties of the system \citep[e.g.,][]{1963P&SS...11..901C, 2015GeoRL..42.9001C, 2017A&A...605L...7L, 2017GeoRL..4411706K, 2018A&A...620A.147B}. For this reason, when observing the upper atmospheres of exoplanets, the transit geometry can have important effects on the interpretation of the data. For example, if a transiting planet has a nonzero impact parameter, a large portion of its exosphere may not transit and thus not contribute to the observed in-transit absorption. Furthermore, a subtler effect in time-series analyses of transmission spectroscopy is the dilution of a planetary absorption signal when the data are co-added in phase space. Since upper atmospheres are extended, the in-transit absorption is variable with time. This variability dilutes the in-transit absorption because time series of transmission spectra are co-added in phase space to improve the signal-to-noise ratio of the combined transmission spectrum \citep[e.g.,][]{2015A&A...577A..62W}.

There are currently no publicly available tools to predict and interpret metastable He transmission spectroscopy. Considering that there is a broad community interest in these observations, we developed {\tt p-winds}, an open-source, fully documented Python implementation of the one-dimensional, isothermal Parker wind\footnote{\footnotesize{In this manuscript, we use the terms "wind" and "outflow" interchangeably, but the first should not be confused with horizontal winds in the lower atmosphere.}} description \citep{1958ApJ...128..664P}, to model exoplanet atmospheres. This code is timely because many data sets used to study metastable He spectroscopy have recently become public. Furthermore, an open-source implementation allows for an independent verification of results as well as community contributions to the code. {\tt p-winds} implements limb darkening and a ray-tracing algorithm that allows the user to change the transit geometry (namely the transit impact parameter and phase in relation to mid-transit).

In this manuscript we describe the overarching implementation of {\tt p-winds}, discuss the design decisions, and illustrate the usage of the code. In Sect. \ref{sect:methods} we describe the several modules implemented in the code to forward model the metastable He signature in a transiting exoplanet. In Sect. \ref{sect:results} we present case studies of the warm Neptunes HAT-P-11~b and GJ~436~b and their corresponding atmospheric escape rates retrieved by fitting {\tt p-winds} models to observations.\ Finally, in Sect. \ref{sect:conclusions} we discuss the conclusions of this work.

\section{Methods}\label{sect:methods}

The code {\tt p-winds} is largely based on the formulations of \citet{2018ApJ...855L..11O} and \citet{2020A&A...636A..13L}. In its current version, the code has four core modules (and two support modules) to model the upper atmospheres and ionization balance of H and He around planetary bodies, which we describe below. In principle, these modules can be used independently of one another depending on the objective of the user. The code to reproduce the examples shown in this section can be obtained via the {\tt p-winds} documentation.

\subsection{The {\tt parker} module}\label{sect:parker}

The {\tt parker} module calculates the structure of the upper atmosphere following the theoretical description of the solar wind by \citet{1958ApJ...128..664P}. In this model, a steady-state, spherically symmetric outflow follows the equation of mass conservation:

\begin{equation}
    \dot{m} = 4 \pi r^2 \rho(r)v(r) \mathrm{,}
\end{equation}where $\dot{m}$ is the mass loss rate, $r$ is the radius, $\rho$ is the gas density, and $v$ is the outflow velocity. This model also follows the steady-state momentum equation:

\begin{equation}
    v \frac{dv}{dr} + \frac{1}{\rho}\frac{dp}{dr} + \frac{G M_\mathrm{pl}}{r^2} = 0 \mathrm{,}
\end{equation}where $G$ is the gravitational constant, $p$ is the thermal pressure and $M_\mathrm{pl}$ is the planetary mass. The isothermal Parker solar wind model assumes that the outflow is completely ionized, yielding a constant mean molecular weight, $\mu$, and consequently a constant sound speed, $v_\mathrm{s}$, as a function of radial distance. This allows for a significant simplification of the problem. However, as pointed out by \citet{2020A&A...636A..13L}, the upper atmosphere of a hot planet differs from the solar wind in that $\mu(r)$ is not necessarily constant with radial distance. But if we assume that the ratio $T(r) / \mu(r)$ is constant over $r$, then the assumption of a constant sound speed profile still holds. If we let $\bar{\mu}$ be the value of mean molecular weight corresponding to a given temperature $T_0$, then the constant sound speed is calculated as

\begin{equation}
    v_\mathrm{s} = \sqrt{\frac{k T_0}{\bar{\mu}}}\mathrm{.}
\end{equation}According to \citet{2020A&A...636A..13L}, the temperature $T_0$ in this approach corresponds to roughly the maximum of the temperature profile obtained by more comprehensive, self-consistent models (see Sect. 3.1 in their manuscript). As we see in the following paragraphs, we arrive at a similar conclusion in our calculations as well.

We calculate $\mu(r)$ as

\begin{equation}
    \mu(r) = m_{\rm p}\,\frac{1 + 4\,n_{\rm He}/n_{\rm H}}{1 + n_{\rm He} / n_{\rm H} + f_{\rm ion}(r)} \mathrm{, with}\ f_{\rm ion} = \frac{n_{\rm H^+}}{n_{\rm H}} \mathrm{,}
 \end{equation}where $m_\mathrm{p}$ is the mass of a proton and $n_\mathrm{X}$ is the number density of species X. For clarity, we note that $n_{\rm H} = n_{\rm H^0} + n_{\rm H^+}$ and $n_{\rm He} = n_{\rm He\,1^1S}~{\rm (singlet)} + n_{\rm He\,2^3S}~{\rm (triplet)} + n_{\rm He^+}$. We assume that the electrons coming from He ionization do not significantly contribute to changes in $\mu$. According to \citet{2018ApJ...855L..11O}, who also make this same assumption, including electrons from He ionization increases their number density by up to $\sim$10\%.

 For a given temperature $T_0$, the corresponding average mean molecular weight, $\bar{\mu}$, is calculated as in Eq. A.3 of \citet{2020A&A...636A..13L}:

 \begin{equation}
    \bar{\mu} = \frac{GM_{\rm pl} \int \mu(r)\frac{dr}{r^2} + \int \mu(r) v(r) dv + kT_0 \int \mu(r)d(1/\mu)}{GM_{\rm pl} \int \frac{dr}{r^2} + \int v(r) dv + kT_0 \int d(1/\mu)} \mathrm{.}
 \end{equation}

It is convenient to normalize the radii, velocities and densities to, respectively, the radius at the sonic point ($r_\mathrm{s}$), the constant speed of sound, and the density at the sonic point ($\rho_\mathrm{s}$). Based on the formulation of \citet{1999isw..book.....L}, the resulting equations describing the radial profiles of velocity and density are

\begin{equation}\label{v_eq}
    \tilde{v}(r) \exp \left[ \frac{-\tilde{v}(r)^2}{2} \right] = \left( \frac{1}{\tilde{r}} \right)^2 \exp \left[ -\frac{2}{\tilde{r}} + \frac{3}{2} \right]\ \mathrm{and}
\end{equation}

\begin{equation}\label{rho_eq}
    \tilde{\rho}(r) = \exp \left[ \frac{2}{\tilde{r}} - \frac{3}{2} - \frac{\tilde{v}^2}{2} \right] \mathrm{,}
\end{equation}where $\tilde{r}$, $\tilde{v}$, and $\tilde{\rho}$ are the normalized radial distance, velocity, and density, respectively.

Calculating the structure of the upper atmosphere requires as input: the planetary parameters and the stellar spectrum from X-rays to ultraviolet (XUV) impinging at the top of the atmosphere, as well as values for the atmospheric temperature and escape rate; the latter two are free parameters in the Parker wind model. Equation \ref{v_eq} is transcendental and requires a numerical approach to determine its solutions. To this end, we utilize a Newton-Raphson method implemented in {\tt scipy.optimize}, which requires an initial guess for the optimization. Equation \ref{v_eq} has many solutions, but we are only interested in the solution that represents an escaping atmosphere (i.e., a transonic solution). In order to guarantee we converge to the correct solution, we enforce that the initial guess is below (above) the speed of sound when calculating the velocities below (above) the sonic point. The end product of the {\tt parker} module is the structure of the upper atmosphere from Eqs. \ref{v_eq} and \ref{rho_eq}.

\begin{figure}
\centering
\includegraphics[width=0.9\hsize]{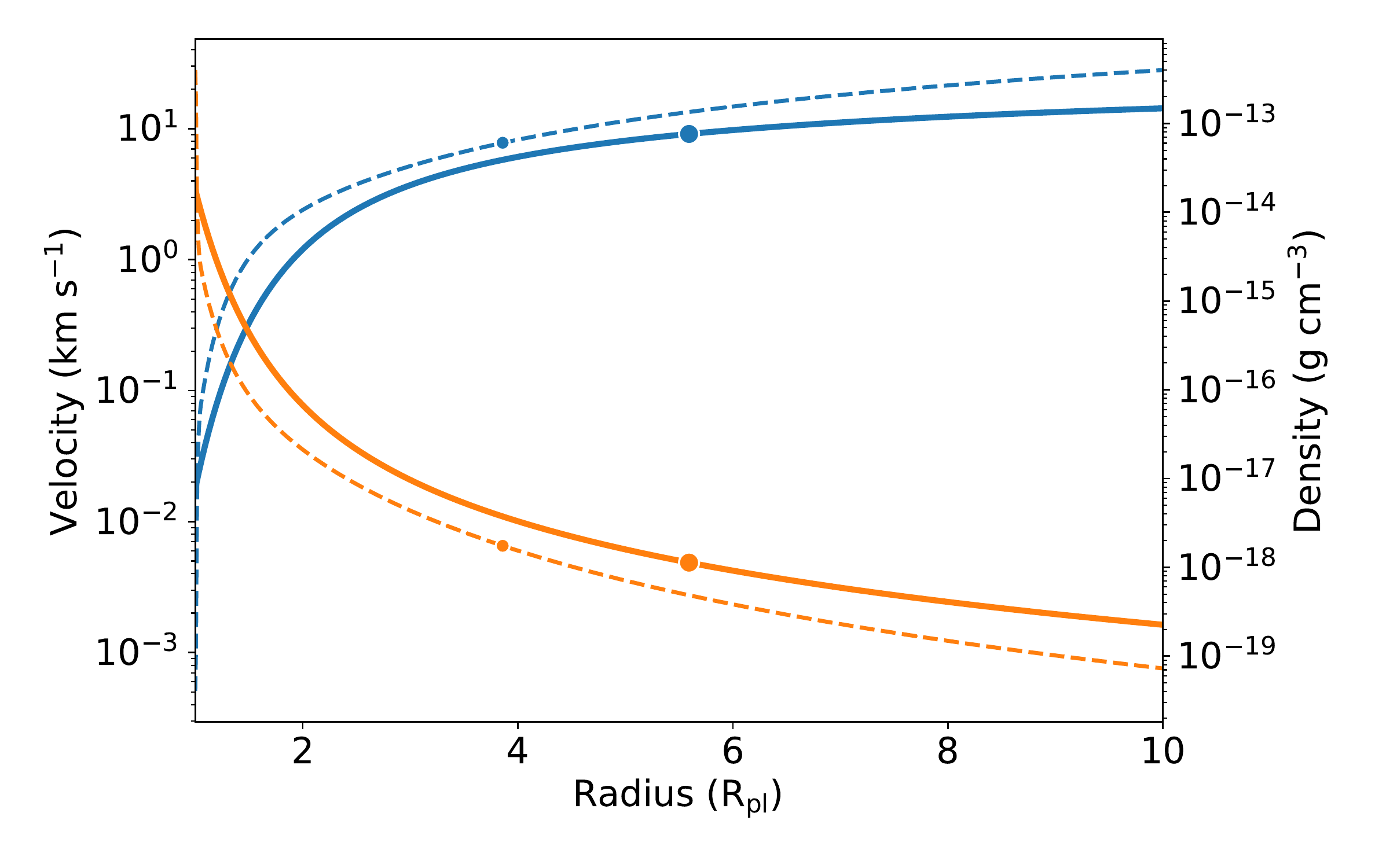}
\caption{One-dimensional structure of the upper atmosphere of the hot Jupiter HD~209458~b computed with {\tt p-winds} (continuous curves). Velocities are shown in blue and densities in orange. For comparison, we also plot a model for the same planet computed self-consistently using the formulation of \citet{2019MNRAS.490.3760A} as dashed curves. The circles mark the sonic point.}
\label{fig:structure_hd209}
\end{figure}

Since the structure of the upper atmosphere (densities and velocities) and the H ionization fraction are interdependent of each other, the code performs a loop that iteratively calculates all of these one-dimensional profiles until convergence is achieved (see Sect. \ref{sect:hydrogen}). As an example, we calculated the structure the hot Jupiter HD~209458~b and the result is shown in Fig. \ref{fig:structure_hd209}. We compare the structure computed with {\tt p-winds} (continuous curves) with a one-dimensional model of the same planet calculated self-consistently using the formulation of \citet[][dashed curves, assuming 90\% H and 10\% He]{2019MNRAS.490.3760A}. In order to be comparable, the {\tt p-winds} model was computed using the same mass loss rate, composition, and a monochromatic ionizing flux (energy $> 13.6$~eV) of 450~erg\,s$^{-1}$\,cm$^{-2}$ with an average photon energy of 20 eV. The isothermal Parker wind predicts a similar structure as the self-consistent model when we assume a $T_0$ corresponding to the maximum temperature from the latter. This is the same result that \citet{2020A&A...636A..13L} obtains in their description.

Naturally, a one-dimensional model does not capture outflow asymmetries that are sometimes observed in Lyman-$\alpha$ transit spectroscopy \citep[e.g.,][]{2010A&A...514A..72L, 2017A&A...605L...7L, 2018A&A...620A.147B}. More complex, three-dimensional models are necessary to completely describe these features \citep[e.g.,][]{2016A&A...591A.121B, 2021MNRAS.501.4383V, 2021ApJ...914...98W, 2021ApJ...914...99W, 2019A&A...623A..58A, 2021arXiv210707534M}. Simple one-dimensional models are nevertheless capable of retrieving atmospheric escape parameters with the assumption that the mass loss process takes place spherically and homogeneously throughout the surface of the planet \citep[e.g.,][]{2020A&A...636A..13L, 2021A&A...647A.129L}. Models that are faster to calculate are also useful when there is a need to explore a large parameter space, which is what we discuss in Sect. \ref{sect:results} and in an upcoming manuscript \citep{Vissapragada2021}.

\subsection{The {\tt hydrogen} module}\label{sect:hydrogen}

The {\tt hydrogen} module calculates the steady-state distribution of neutral and ionized H in the upper atmosphere. The quantity of interest here is $f_\mathrm{ion}$, whose radial profile is obtained by calculating the steady-state balance between advection and source-sink terms for H ions. In this case, the source is (photo-) ionization by high-energy photons, and the sink is recombination into neutral atoms. This radial distribution can be calculated with the following differential equation \citep[see Sect. 3.2 in][]{2018ApJ...855L..11O}:

\begin{equation}\label{ss_H}
v(r)\,\frac{d f_\mathrm{ion}}{d r} = (1 - f_\mathrm{ion})\,\Phi(r) - n_{\rm H}(r)\,f_\mathrm{ion}^2\,\alpha_\mathrm{rec} \mathrm{,}
\end{equation}where $n_{\rm H}(r) = x\,\rho(r) / [(x + 4y)\,m_p]$, with $x$ being the H atoms number fraction in the outflow, $y = 1 - x$  is the He atoms number fraction, and $\Phi$ is the photoionization rate:

\begin{equation}
    \Phi(r) = \int^{\lambda_0}_{0} \frac{\lambda}{hc}\,f_\lambda\,\sigma_\lambda\,e^{-\tau_\lambda(r)}\,d\lambda \mathrm{,}
\end{equation}where $\lambda_0$ is the wavelength corresponding to the ionization energy of H (911.65 \AA) and $f_\lambda$ is the flux density (in units of energy $\cdot$ time$^{-1} \cdot$ area$^{-1} \cdot$ wavelength$^{-1}$) arriving at the top of the atmosphere. $\sigma_\lambda$ is the photoionization cross section, which we calculate in the support module {\tt microphysics}, following Eq. 10 in \citet{2018ApJ...855L..11O}, which is based on \citet{2006agna.book.....O}. The optical depth of neutral H is given by

\begin{equation}\label{tau_eq}
    \tau_{\lambda,\,H^0}(r) = \int_{r}^{\infty} \sigma_\lambda\,n_{H^0}(r)\,dr = \frac{x\,\sigma_{\lambda}}{(x + 4y)\,m_p}\int_{r}^{\infty} (1 - f_{\rm ion})\,\rho(r)\,dr
.\end{equation}

The velocities $v$ and densities $\rho$ are calculated using the module {\tt parker}. $\alpha_{\rm rec}$ is the case-B H recombination rate at a given temperature \citep{2006agna.book.....O, 2015ApJ...808..173T}, calculated as

\begin{equation}
    \alpha_{\rm rec} = 2.59 \times 10^{-13} \left(\frac{T_0}{10^4}\right)^{-0.7}~\mathrm{cm}^3~\mathrm{s}^{-1}\mathrm{.}
\end{equation}

As seen in Eq. \ref{tau_eq}, $\tau_\lambda$ depends on $f_\mathrm{ion}$, which is what we want to calculate in the first place. However, the optical depth depends more strongly on the densities of H than the ion fraction. So instead of solving a system of coupled nonlinear differential equations, a first solution can be achieved by assuming that the whole atmosphere is neutral at first. Later, we relax this assumption by recalculating the $\tau_\lambda$ and $f_{\rm ion}$ profiles iteratively until convergence is achieved (the user can define the convergence criterion). 

\begin{figure}
\centering
\includegraphics[width=0.9\hsize]{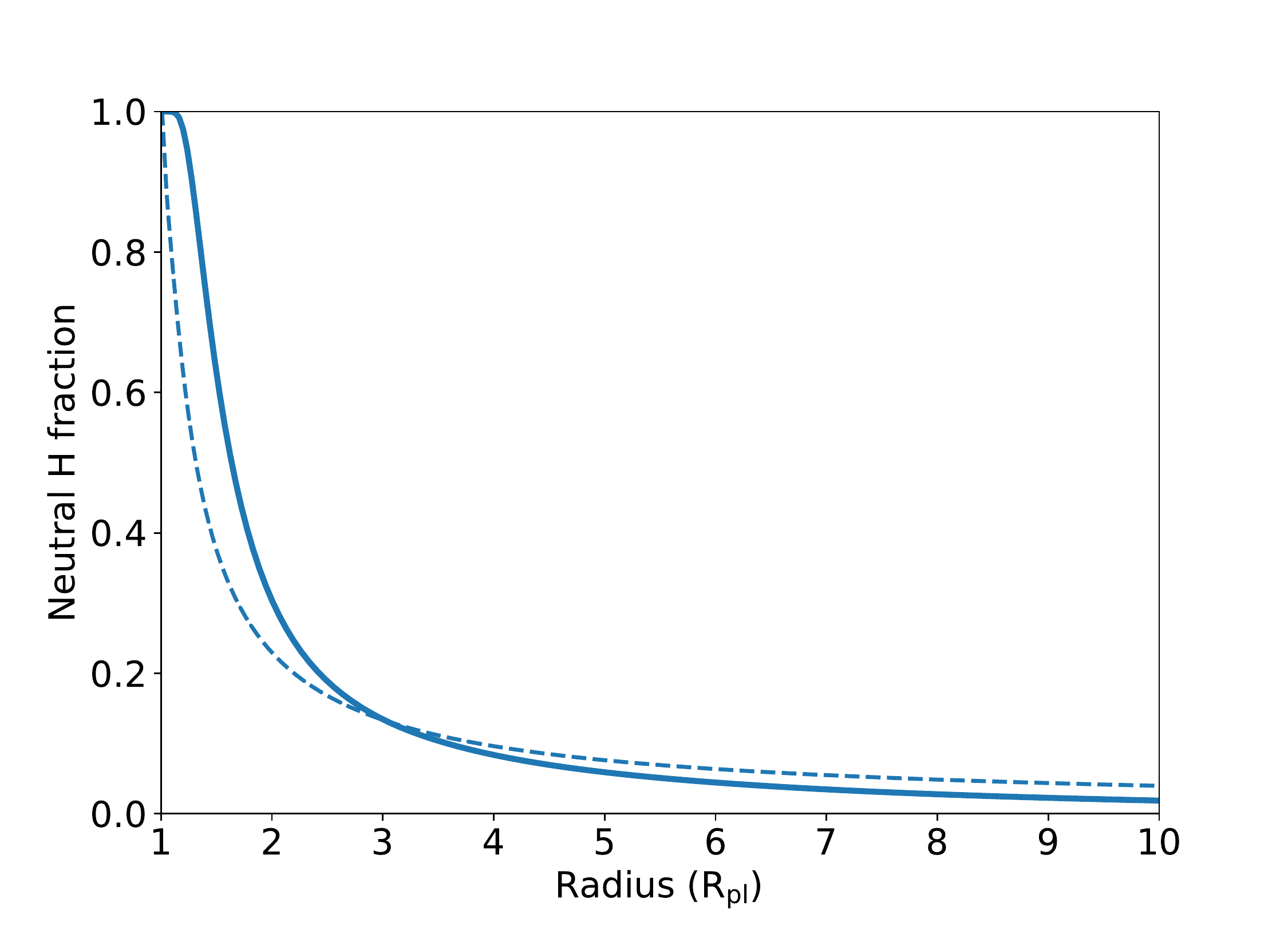}
\caption{Neutral H atom fraction in the upper atmosphere of the hot Jupiter HD~209458~b computed with {\tt p-winds} for the same setup from Sect. \ref{sect:parker} (continuous curve). We also show the neutral fraction calculated with a self-consistent escape model for comparison (dashed curve).}
\label{fig:hion_hd209}
\end{figure}

We solve Eq. \ref{ss_H} using {\tt solve\_ivp}, an explicit Runge-Kutta integrator of hybrid 4th and 5th orders implemented in \texttt{scipy.integrate}. The user inputs an initial guess for $f_{\rm ion}$ at the innermost layer of the upper atmosphere. The code also takes as input the stellar host spectrum arriving at the planet, or the monochromatic flux between 0 and 911.65 \AA, and the planetary parameters. The solution for the H distribution in 500 points including the relaxation takes approximately 400~ms on a CPU with frequency 3.1~GHz and four computing threads. Continuing the example for HD~209458~b from Sect. \ref{sect:parker}, we calculated the ion and neutral fraction of H in the upper atmosphere, and the resulting distribution is shown in Fig. \ref{fig:hion_hd209} (continuous curve). We compare this result with the ion fraction calculated with the self-consistent escape model from Sect. \ref{sect:parker} (dashed curve); in order to be comparable, both models are calculated assuming an impinging XUV monochromatic flux of 450~erg~s$^{-1}$~cm$^{-2}$. The {\tt p-winds} model overpredicts the ion fraction by a factor of a few when compared to the self-consistent model, likely because of the larger densities (see Fig. \ref{fig:structure_hd209}), which increase the optical depth of the atmosphere to ionizing irradiation.

\subsection{The {\tt helium} module}

The {\tt helium} module calculates the steady-state distribution of neutral singlet, neutral triplet, and ionized He in the upper atmosphere. The quantities of interest here are $f_1 = n_{\rm He\,1^1S} / n_{\rm He}$ and $f_3 = n_{\rm He\,2^3S} / n_{\rm He}$. The radial profiles $df_1/dr$ and $df_3/dr$ are described by a coupled system of differential equations with source and sink terms:

\begin{equation}\label{eq:he_dist}
    \begin{cases}
     v(r)\,d f_1 / d r = \mathrm{sources}_1 + \mathrm{sinks}_1 \\
     v(r)\,d f_3 / d r = \mathrm{sources}_3 + \mathrm{sinks}_3 
    \end{cases}\mathrm{.}
\end{equation}We refer the reader to \citet{2018ApJ...855L..11O} and Table 2 of \citet{2020A&A...636A..13L} for detailed equations of all the source and sink terms for He\footnote{\footnotesize{We note that, in Table 2 of \citet{2020A&A...636A..13L}, the units for the recombination and collisional processes are cm$^3$\,s$^{-1}$, and not cm$^{-3}$\,s$^{-1}$ as the authors list in their manuscript.}}. In our code we do include the He charge exchange terms pointed out by \citet{2020A&A...636A..13L}.

We assume that the He ionization and the excited helium triplet do not significantly change the structure of the upper atmosphere. This allows us to decouple the {\tt helium} module from the {\tt parker} and {\tt hydrogen} modules. This is advantageous because the user can enter as input a H structure that was calculated by more complex and self-consistent models than isothermal Parker winds ones. It is important, however, that these models do include He in their calculation of the structure in order to produce consistent results for the metastable He distribution.

The procedure to solve the distribution of He (Eq. \ref{eq:he_dist}) is similar to that for H. The user inputs an initial guess for $f_1$ and $f_3$ at the innermost atmospheric layer, the stellar host spectrum from 0 to 2600~\AA\ (or monochromatic fluxes in the bands 0-1200~\AA\ and 1200-2600~\AA), the structure of the upper atmosphere (profiles of density and velocity), and the planetary parameters. It is important to emphasize that neutral H also contributes to the optical depth between wavelengths 0-911~\AA, attenuating the amount of high-energy flux that ionizes and populates the He levels. The code takes this contribution into account, as in \citet{2018ApJ...855L..11O}.

Initially, the code assumes that the entire upper atmosphere has constant $f_1$ and $f_3$, and then a first solution is obtained using {\tt odeint}\footnote{\footnotesize{When calculating the steady-state distribution of He, we opted to use {\tt odeint} instead of {\tt solve\_ivp} in this case because the first is more stable and 2.6 times faster than the second. {\tt solve\_ivp} is faster than {\tt odeint} when solving the distribution of H.}}, a Python wrapper for the {\tt LSODA} solver from the Fortran library {\tt odepack} implemented in {\tt scipy.integrate}. This solution is then relaxed by updating the optical depths, $f_1$ and $f_3$ iteratively until convergence is achieved. The solution can, however, become numerically unstable for large density gradients, which can sometimes happen near the $R = 1$~R$_{\rm pl}$. A practical work-around is to establish a cutoff near $1$~R$_{\rm pl}$ that removes this large density gradient and ignore this layer of the atmosphere in the modeling. However, the user should be aware that this solution could affect the interpretation of more compressed thermospheres, such as that of HD~189733~b \citep{2021A&A...647A.129L}; we have not yet attempted to model this planet with \texttt{p-winds}, and leave this for future work.

\begin{figure}
\centering
\includegraphics[width=0.9\hsize]{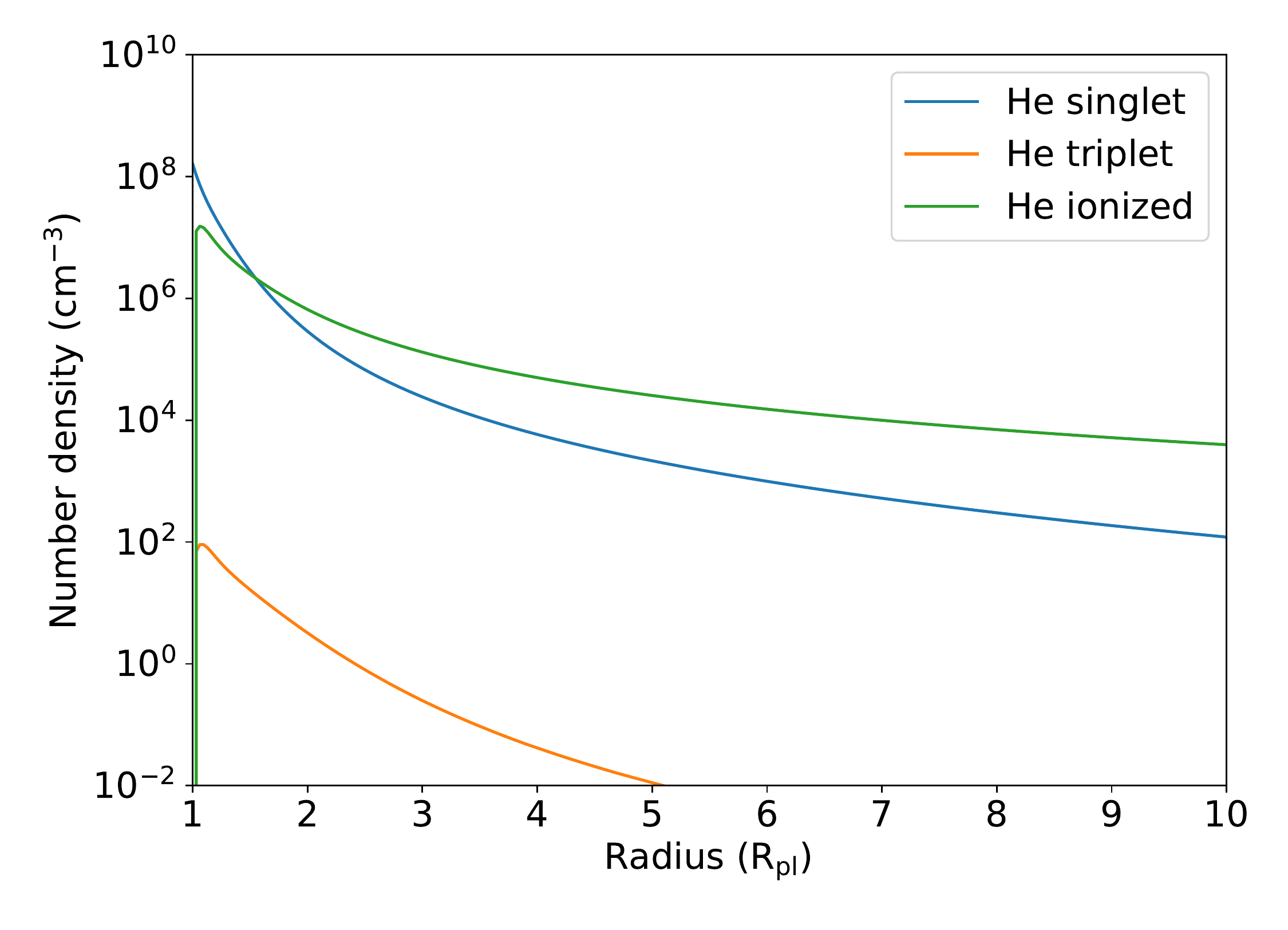}
\caption{Distribution of He in the upper atmosphere of HD~209458~b calculated with {\tt p-winds} assuming the same input parameters as \citet{2018ApJ...855L..11O}.}
\label{fig:he_dist_hd209}
\end{figure}

We show the distribution of He in the upper atmosphere of HD~209458~b calculated as described above in Fig. \ref{fig:he_dist_hd209}.  For comparison purposes, this time we assumed a model with the same input parameters as the one described in \citet[][namely an escape rate of $8 \times 10^{10}$~g~s$^{-1}$, a temperature of 9000~K, a H fraction of 0.9, and a solar irradiating spectrum]{2018ApJ...855L..11O}. Our results match the models of \citeauthor{2018ApJ...855L..11O}, as seen in Fig. 3 of their publication. The solution for the He distribution in 500 points including the relaxation takes approximately 2.5~s on a CPU with frequency 3.1~GHz and four computing threads. This is the main computational bottleneck of the {\tt p-winds} code.

\subsection{The {\tt transit} module}

The {\tt transit} module has two independent functions that can be used to calculate the spectral signatures of the upper atmosphere in transmission. The first function, {\tt draw\_transit}, calculates two-dimensional intensity maps containing the host star and a transiting planet at a user-defined phase and impact parameter. The one-dimensional profiles of metastable He volumetric densities are required to calculate the two-dimensional array of column densities mapped to the same geometry as the transit. The output intensity map is normalized in a way that the disk-averaged stellar intensity is 1.0 when the planet is out of transit. Optionally, the user can also input a limb-darkening law.

The most important function in this module is {\tt radiative\_transfer\_2d}, which, as the name implies, calculates the in-transit absorption spectrum caused by the opaque disk of the planet and its upper atmosphere. In each cell of the two-dimensional transit mapped by the $ij$ indexes, the resulting attenuated intensity $I_{ij}(\nu)$\footnote{\footnotesize{The radiative transfer routine uses input in wavelength space, but the actual calculations are performed in frequency space for code clarity and brevity. The grid size is defined by the user.}} of the stellar light caused by absorption of He in the upper atmosphere is given by

\begin{equation}\label{eq:rad_transf}
    I_{ij}(\nu) = I_{ij,\,0}(\nu) \exp{(-\tau_{ij,\,{\rm He}})} \mathrm{,}
\end{equation}where $I_{ij,\,0}(\nu)$ is the intensity emerging from the host star before filtering through the atmosphere and $\tau_{ij,\,{\rm He}}$ is the optical depth due to metastable He. $I_{ij}(\nu)$ is set to zero in the cells corresponding to the opaque disk of the planet. From here onward, we drop the $ij$ indexes for the sake of brevity, but the reader should implicitly assume that the radiative transfer is carried out cell-by-cell in the transit map. Formally, the optical depth is given by

\begin{equation}\label{eq:opt_depth}
    \tau_{\rm He} = \int_{-R_{\rm atm}}^{R_{\rm atm}} \varphi_\nu(z)\,\sigma_{\rm He}\,n_{\rm He}(z)\, dz \mathrm{,}
\end{equation}where $\varphi_\nu$ is the Voigt profile and $\sigma_{\rm He}$ is the cross section of metastable helium lines near 1.083~$\mu$m. Following \citet{2018ApJ...855L..11O} \citep[see also, e.g.,][]{2019MNRAS.490.3760A}, the He cross section is calculated as

\begin{equation}
    \sigma_{\rm He} = \frac{\pi e^2}{m_e c} f \mathrm{,}
\end{equation}where $f$ is the oscillator strength of the transition, $e$ is the electron charge, and $m_e$ is the electron mass. This formula is only valid in the Gaussian-cgs unit system, where $e$ is given in units of esu or statC (see, for example, \citealt{2010ApJ...723..116K} for a formula that can be used in other unit systems). 

The Voigt profile $\varphi_\nu$ is calculated using the {\tt voigt\_profile} implementation of {\tt scipy.special}, which takes three parameters: the bulk velocity shift $v_{\rm bulk}$ of the profile in relation to the rest wavelength, the standard deviation $\alpha$ of the Gaussian (in our case Doppler) term, and the Lorentzian half width at half maximum (HWHM). Similar to \citet{2020A&A...636A..13L}, the Gaussian width $\alpha$ is calculated as

\begin{equation}\label{eq:gaussian_broadening}
    \alpha = \frac{\nu_0}{c} \sqrt{\frac{2\,k_B T}{m_{\rm He}}} \mathrm{,}
\end{equation}where $m_{\rm He}$ is the mass of a He atom, $T$ is the temperature of the gas, $\nu_0$ is the central frequency of the transition. The Lorentzian HWHM is $\gamma = A_{ij} / 4\pi$, where $A_{ij}$ is the Einstein coefficient of the transition. We took the properties of the metastable He transitions near 1.083~$\mu$m from the National Institute of Standards and Technology (NIST) database\footnote{\footnotesize{\url{https://www.nist.gov/pml/atomic-spectra-database}.}}, and list them in Table \ref{triplet_properties}.

\begin{table}
\caption{Spectral line properties of the metastable He triplet in the near-infrared.}
\label{triplet_properties}
\centering
\begin{tabular}{cccc}
\hline\hline
 \multirow{2}{*}{Upper level $J$} & $\lambda_0$ & $A_{ij}$ & \multirow{2}{*}{$f$} \\
  & (nm, in air) & (s$^{-1}$) & \\
  \hline
 $0$ & $1\,082.909$ & $1.0216 \times 10^7$ & $5.9902 \times 10^{-2}$ \\
 $1$ & $1\,083.025$ & $1.0216 \times 10^7$ & $1.7974 \times 10^{-1}$ \\
 $2$ & $1\,083.034$ & $1.0216 \times 10^7$ & $2.9958 \times 10^{-1}$ \\
\hline
\end{tabular}
\end{table}

\subsubsection{Line broadening by the planetary outflow}

In reality, $\varphi_\nu$ depends on the three-dimensional position in relation to the planet because each position has a different line-of-sight velocity, which broadens the absorption line. Thus, the formal calculation of the Voigt profile is performed for each pencil of light between the star and the observer. In a given position $z$ along the pencil, the line-of-sight velocity, $v_{\rm LOS}$, as a function of distance, $r$, from the planet center is calculated using the formulation of \citet{2020A&A...633A..86S}:

\begin{equation}
    |v_{\rm LOS}(r)| = |v_{\rm ver}| \frac{z^2}{\sqrt{r^2 + z^2}} \mathrm{,}
\end{equation}where $v_{\rm ver}$ is the outflow velocity obtained from the Parker wind model. 

This calculation has to be performed for three spectral lines, and it adds an extra dimension for wavelength. For these reasons, the formal calculation of $\varphi_\nu$ taking into account all the four dimensions is computationally costly. In order to accelerate the radiative transfer, instead of calculating the Parker wind broadening in full dimensionality, we can optionally assume that it contributes to the Gaussian broadening term of the Voigt profile uniformly through the line of sight. With the dependence on the $z$ axis dropped, we can remove $\varphi_\nu$ from the integrand in Eq. \ref{eq:opt_depth}, yielding the approximation

\begin{equation}\label{eq:rad_transf_approx}
    \tau_{\rm He} \simeq \varphi_\nu\,\sigma_{\rm He} \int_{-R_{\rm atm}}^{R_{\rm atm}} n_{\rm He}(z)\, dz = \varphi_\nu\,\sigma_{\rm He}\,\eta_{\rm He} \mathrm{,}
\end{equation}where $\eta_{\rm He}$ is the column density of He. In order to validate this approximation, we need to assume that the Gaussian wind broadening has a constant velocity $v_{w}$ in the line of sight, and add it in quadrature to the square-velocity term of Eq. \ref{eq:gaussian_broadening}, yielding

\begin{equation}\label{eq:alpha_approx}
    \alpha_{\rm approx} = \frac{\nu_0}{c} \sqrt{\frac{2\,k_B T}{m_{\rm He}} + v_{w}^2} \mathrm{.}
\end{equation}The wind-broadening velocity term $v_{w}$ is calculated as the average of $v_{\rm LOS}$ weighted by the metastable He number density:

\begin{equation}\label{eq:vw_average}
    v_{w} = \frac{\int_0^{R_{\rm sim}} v_{\rm LOS}(r)\,n_{\rm He\,2^3S}(r)\,dr}{\int_0^{R_{\rm sim}} n_{\rm He\,2^3S}(r)\,dr} \mathrm{.}
\end{equation}

In this approximation, the user can, optionally, include an additional source of broadening: microturbulence. We implement the same formulation of \citet{2020A&A...636A..13L}:

\begin{equation}\label{eq:vw_turb}
    v_{\rm turb} = \sqrt{5k_B T / (3m_{\rm He})} \mathrm{.}
\end{equation}The turbulence velocity term is added quadratically in Eq. \ref{eq:alpha_approx}.

{\tt p-winds} allows the user to decide on which method to use to calculate the Voigt profile: the formal calculation (Eqs. \ref{eq:rad_transf} and \ref{eq:opt_depth}) or the density-weighted average broadening parameter (Eqs. \ref{eq:rad_transf_approx}, \ref{eq:alpha_approx}, and \ref{eq:vw_average}). The turbulent broadening (Eq. \ref{eq:vw_turb}) can optionally be included at the discretion of the user. This is done with the optional parameters {\tt wind\_broadening\_method} and {\tt turbulence\_broadening} when calling the {\tt radiative\_transfer\_2d} function; the default is the density-weighted average-velocity implementation, which is a good compromise of speed and accuracy.

\begin{figure*}
\centering
\begin{tabular}{cc}
\includegraphics[width=0.47\hsize]{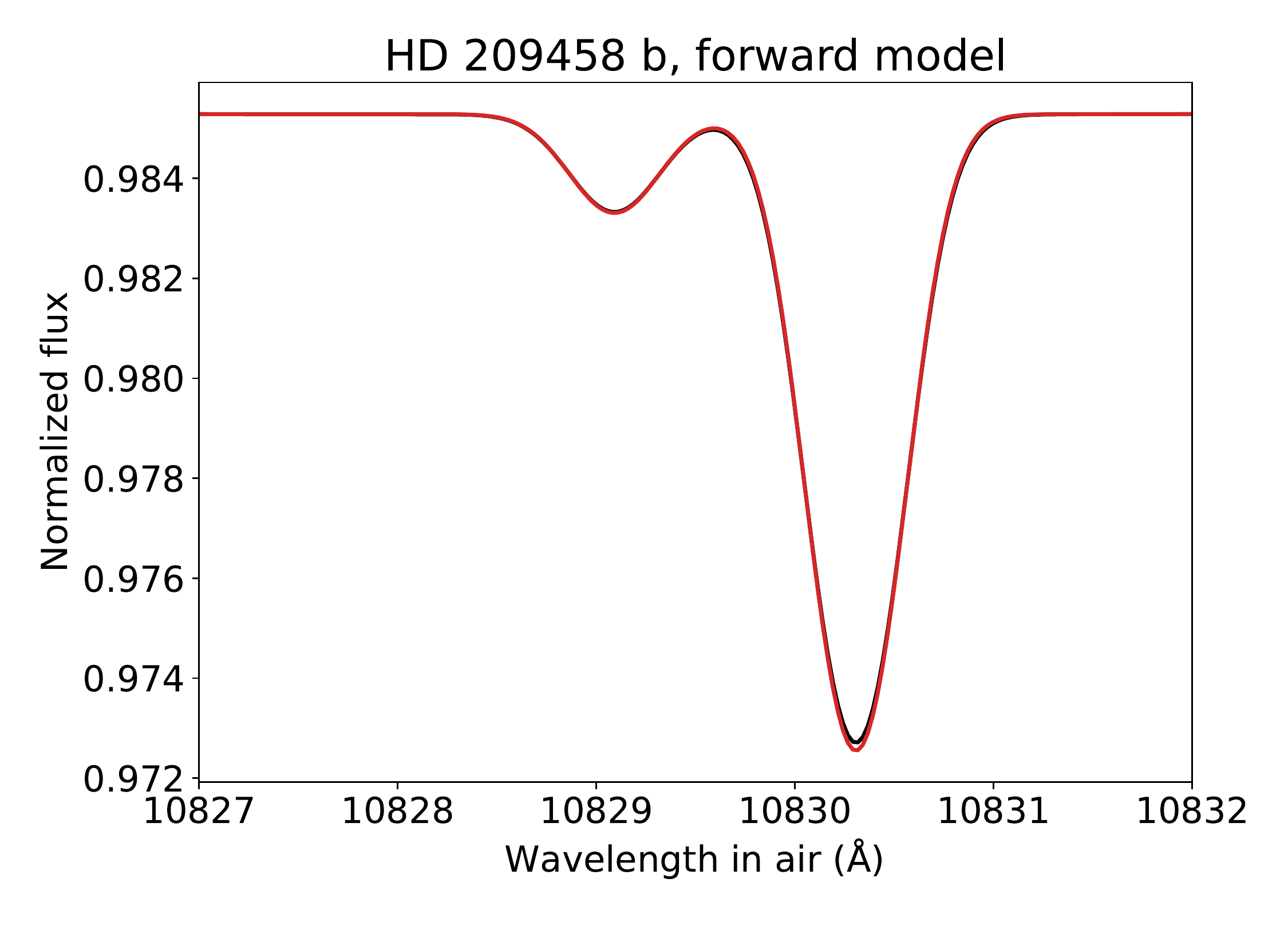} & \includegraphics[width=0.47\hsize]{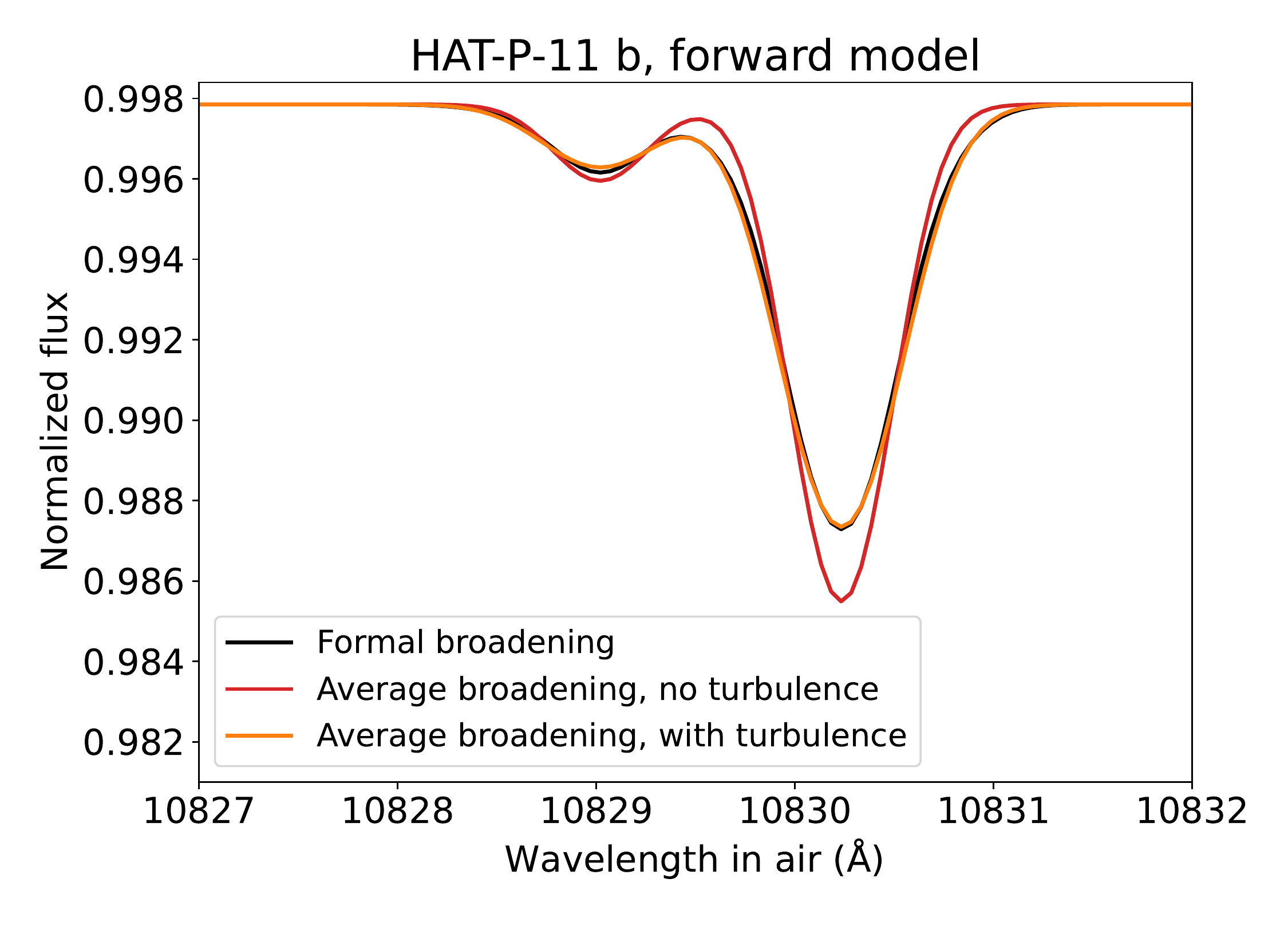}
\end{tabular}
\caption{Comparison of the spectral line broadening in the He triplet lines using two different methods: formal definition of the optical depth (black) and the density-weighted average-velocity broadening (red). In the case of a more extended atmosphere of HAT-P-11~b, a better match between the formal and average methods is obtained when we include turbulence broadening (orange). These are forward models, and we are not yet attempting to fit them to observed signatures.}
\label{fig:profile_comparison}
\end{figure*}

We assess the validity of the assumption we made above by calculating the formal and the average-velocity broadening methods for HD~209458~b and HAT-P-11~b. The first planet has a more compact upper atmosphere and lower outflow velocities than the second by a factor of $\sim$2. In the case of HD~209458~b, the average-velocity method produces an accurate approximation to the formal calculation (see the left panel of Fig. \ref{fig:profile_comparison}). In the case of the more extended atmosphere of HAT-P-11~b, the average-velocity method is accurate when turbulence broadening is also included (right panel of Fig. \ref{fig:profile_comparison}). In both cases, the average method is one order of magnitude faster in computation time than the formal method.

We emphasize that, at this point, we are only producing a forward model and not making an attempt to fit it to actual existing observations of this planet \citep[e.g.,][]{2020A&A...636A..13L}. The results we obtain from the example of HD~209458~b throughout this section, from the Parker wind structure to the predicted metastable He transmission spectrum, are consistent with those obtained by \citet{2018ApJ...855L..11O}. 

\subsubsection{Dilution of the transit signature}

Usually, the absorption of light by upper atmospheres in transiting exoplanets is in the order of several percent or less in a narrow bandpass. Thus, transmission spectroscopy observations sometimes rely on averaging time series in phase-space to build enough signal to noise and produce a detectable signal. However, upper atmospheres are so extended that the in-transit absorption signature is variable with phase, and phase-averaging them dilutes the observed signature. In addition, inhomogeneities in the stellar surface, such as limb darkening, may become important, particularly when fitting (spectro-) photometric light curves.

\begin{figure*}
\centering
\begin{tabular}{cc}
\includegraphics[width=0.47\hsize]{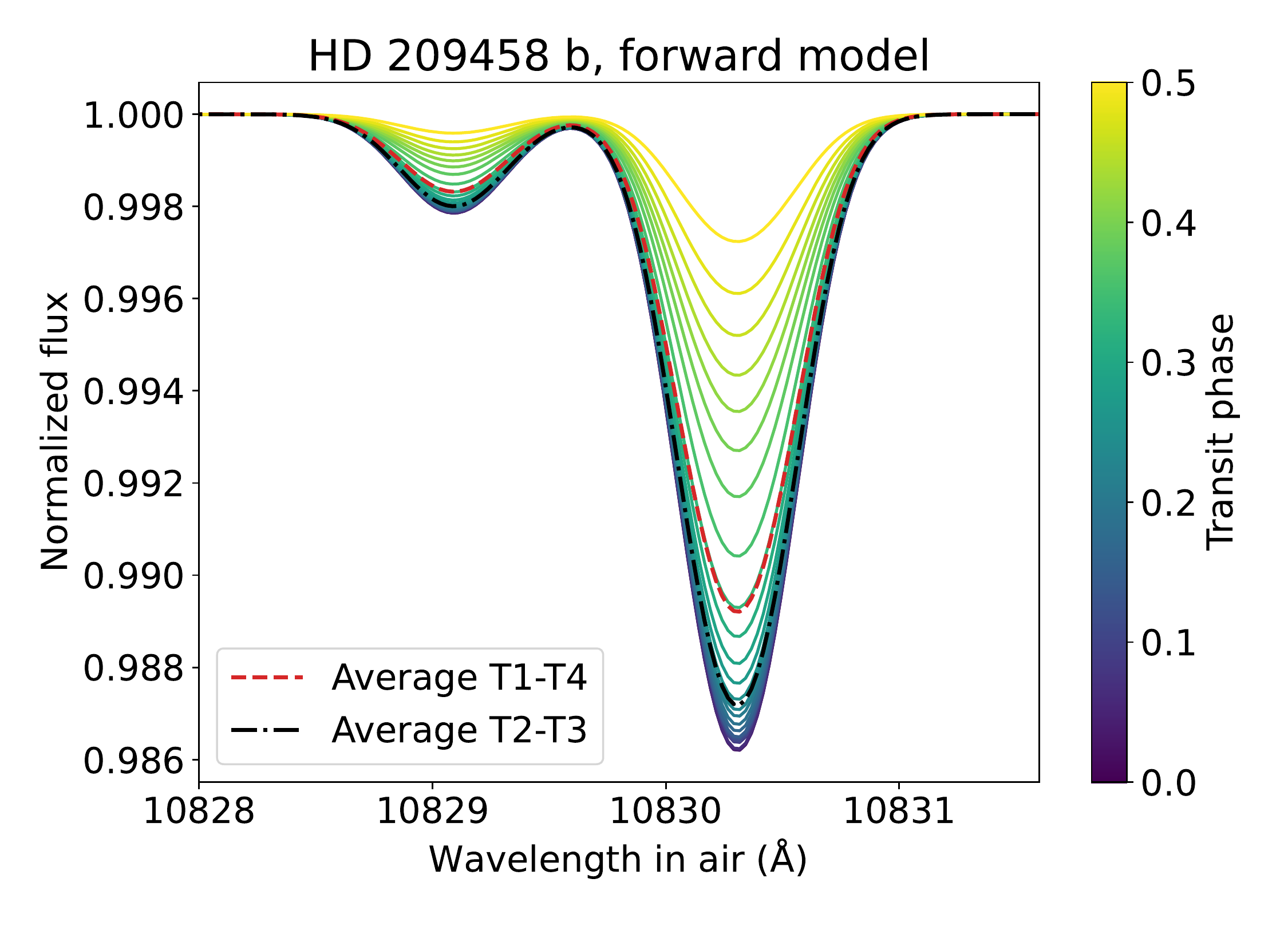} & \includegraphics[width=0.47\hsize]{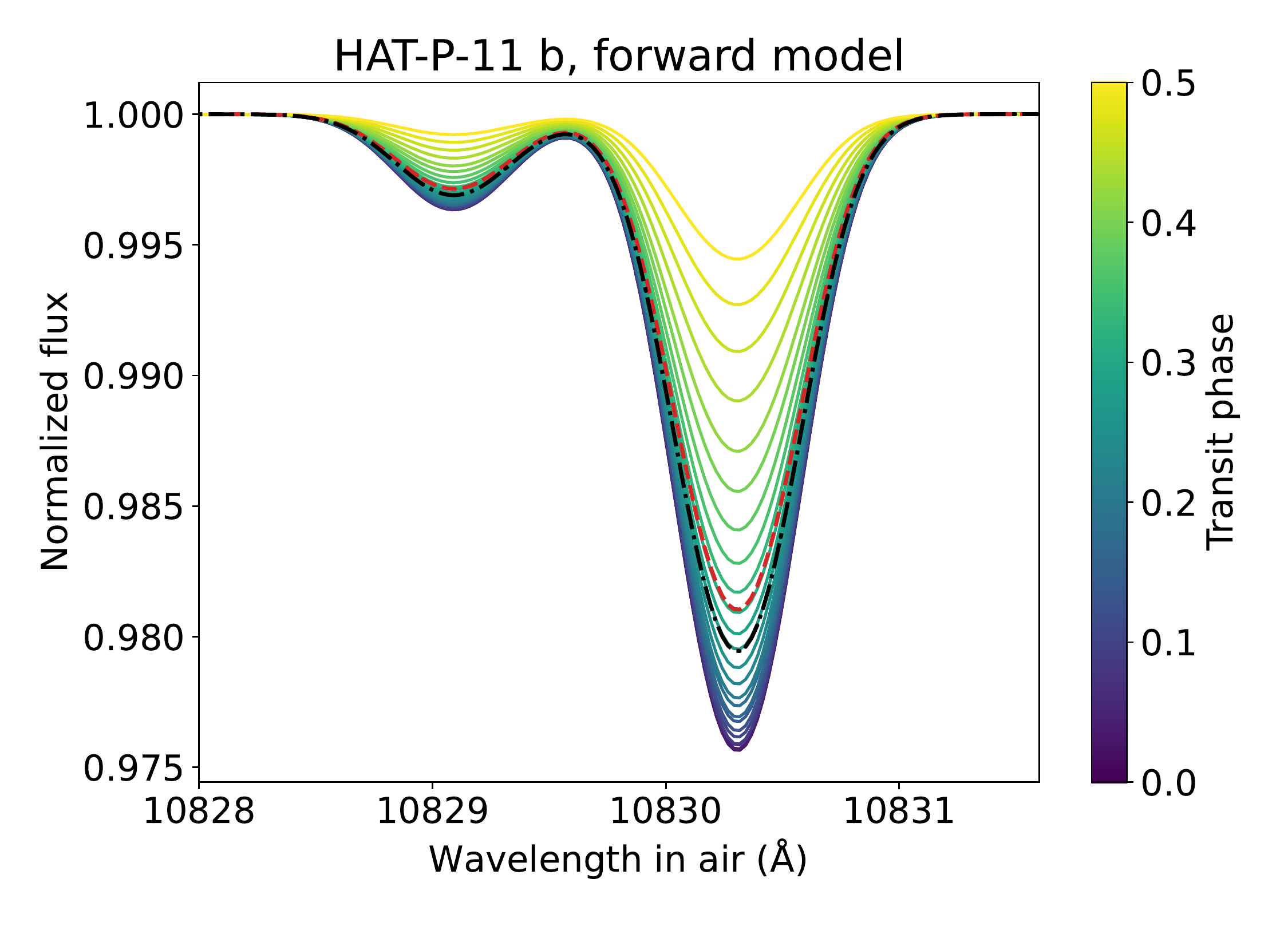} \\
\end{tabular}
\caption{Metastable He transmission spectrum of HD~209458~b (left panel) and HAT-P-11~b (right panel) for uniformly sampled phases, transit impact parameter $b = 0.499$ \textbf{and $b = 0.132$, respectively}, and limb darkening based on the results of \citet{2007ApJ...655..564K} and \citet{2010A&A...510A..21S}. The baseline $(R_{\rm p} / R_{\rm s})^2$ was removed, as in actual ground-based observations. Phase 0.0 and 0.5 represent, respectively, the times of mid-transit and of first (or fourth) contact. The dashed red spectrum is the average of all phases between the first and fourth contact. The dot-dashed black curve is the average between second and third contact. These are forward models, and we are not yet attempting to fit them to observed signatures.}
\label{fig:spec_ts}
\end{figure*}

We illustrate this effect in Fig. \ref{fig:spec_ts}, where we simulated the phase-averaging for both HD~209458~b and HAT-P-11~b. In the case of the hot-Jupiter, the mid-transit spectrum (phase $= 0.0$) is comparable to the spectrum phase-averaged between second and third contacts (T2-T3), but it differs more significantly when the phase-averaging is taken between first and fourth contacts (T1-T4). This is due to a more compact upper atmosphere than HAT-P-11~b, whose extent produces a larger difference between the phase-averaged and the mid-transit spectra. Since the planet-to-star ratio is smaller than HD~209458~b, phase averaging between T1-T4 or T2-T3 does not make as much difference as it does for the hot Jupiter.

Previous one-dimensional descriptions of the metastable He transmission spectrum did not take into account the transit geometry, phase-averaging, and limb darkening. The {\tt p-winds} code allows the user to set the transit impact parameter, phase in relation to the first and fourth contacts, and set a limb-darkening law. To this end, we utilize the auxiliary open-source code {\tt flatstar}\footnote{\footnotesize{The code is freely available at \url{https://flatstar.readthedocs.io}.}} to simulate transit grids (see a brief description in Appendix \ref{app:flatstar}). In the current implementation, this transit grid only allows for circular orbits and it neglects the curvature of the transit chord.

\section{Atmospheric escape retrievals}\label{sect:results}

We further benchmarked the code {\tt p-winds} by performing retrievals of the atmospheric escape rate, temperature, line-of-sight bulk velocity, and the H fraction ($n_{\rm H}/n_{\rm atoms}$) of the warm Neptunes HAT-P-11~b and GJ~436~b. Both planets were observed in transmission spectroscopy by the CARMENES spectrograph (Calar Alto high-Resolution search for M dwarfs with Exoearths with Near-infrared and optical Échelle Spectrographs), but only the first had a strong in-transit signal and the second had only a non-detection. For the latter, we attempted to fit upper or lower limits of the atmospheric escape rate and outflow temperature. The Python algorithms to reproduce our retrievals are freely available online\footnote{\footnotesize{\url{https://zenodo.org/record/4906091}.}}.

\subsection{Fitting the He signature of HAT-P-11~b}\label{retrieval}

The metastable He signature of HAT-P-11~b was measured with the CARMENES spectrograph installed in the 3.5~m telescope at the Calar Alto Observatory \citep{2018Sci...362.1384A}. The transmission spectrum is openly available in the DACE platform\footnote{\footnotesize{\url{https://dace.unige.ch/openData/}}}. The central wavelengths of the metastable He transitions retrieved from the NIST database are listed as measured in air, but the wavelengths of the CARMENES spectrum are in vacuum. We converted the wavelengths of the latter to in-air using the following formula \citep[][IAU standard]{2000ApJS..130..403M}:

\begin{multline}
   \lambda_{\rm air} = \lambda_{\rm vacuum} / n\ {\rm with}\\
    n = 1 + 0.0000834254 + \frac{0.02406147}{130 - s^2} + \frac{0.00015998}{38.9 - s^2}\ {\rm and}\\
    s = 10^4 / \lambda_{\rm vacuum} \mathrm{.}
\end{multline}

In general, we fit three free parameters: the atmospheric escape rate $\dot{m}$, the upper atmosphere temperature $T$, and the bulk line-of-sight velocity $v_{\rm bulk}$ of the upper atmosphere. It is also possible to run fits with additional parameters (such as the H fraction). The fit is performed by maximizing the likelihood $\mathcal{P}$ of a given transmission spectrum model $\mathcal{F}_{\rm model}$ to represent the observed transmission spectrum $\mathcal{F}$. Such a log-likelihood is given by

\begin{equation}\label{eq:likelihood}
    \ln{\mathcal{P}(\mathcal{F} | \lambda, \sigma, \vec{p})} =
    -\frac{1}{2} \sum_n \left[ \frac{\left(\mathcal{F}_n - \mathcal{F}_{\mathrm{model},\,n}\right)^2}{\sigma_n^2} + \ln{\left(2\pi \sigma_n^2 \right)} \right] \mathrm{,}
\end{equation}where $n$ stands for a given bin of the spectrum, $\sigma$ is the uncertainty of the measurement, and $\vec{p}$ is the vector containing the free parameters. To determine the uncertainties of the fit, we use the Markov chain Monte Carlo (MCMC) ensemble sampler {\tt emcee} \citep{2013PASP..125..306F}. The uncertainties we report here represent the confidence interval that encompasses the 16th to the 84th percentile of the posterior distribution of the free parameters.

For HAT-P-11~b, we ran in total four different models: (1) no limb darkening, H number fraction fixed to 0.90; (2) a quadratic limb-darkening law with coefficients $c_1 = 0.63$ and $c_2 = 0.09$ \citep{2010A&A...510A..21S}, H number fraction fixed to 0.90; (3) no limb darkening, H fraction as a free parameter with a uniform prior of [0.80, 0.99]; and (4) same as (1), but using the formal implementation of the radiative transfer instead of the average-velocity broadening approximation. For model (4), instead of a full MCMC, we perform only a maximum-likelihood (Eq. \ref{eq:likelihood}) using the Nelder-Mead algorithm implemented in {\tt scipy.optimize.minimize}. The reason is because we simply want to assess the accuracy of the average-velocity broadening approximation for HAT-P-11~b in comparison to the formal, more computationally costly radiative transfer. HAT-P-11 does not have a full high-energy spectrum measurement, so we use the spectrum of a similar star from the MUSCLES Treasury Survey\footnote{\footnotesize{Available at \url{https://archive.stsci.edu/prepds/muscles/}.}} \citep{2016ApJ...820...89F, 2016ApJ...824..101Y, 2016ApJ...824..102L} as a proxy. We chose the star HD~40307, which has similar effective temperature, mass, radius, and surface gravity as HAT-P-11.

\begin{figure}
\centering
\includegraphics[width=0.9\hsize]{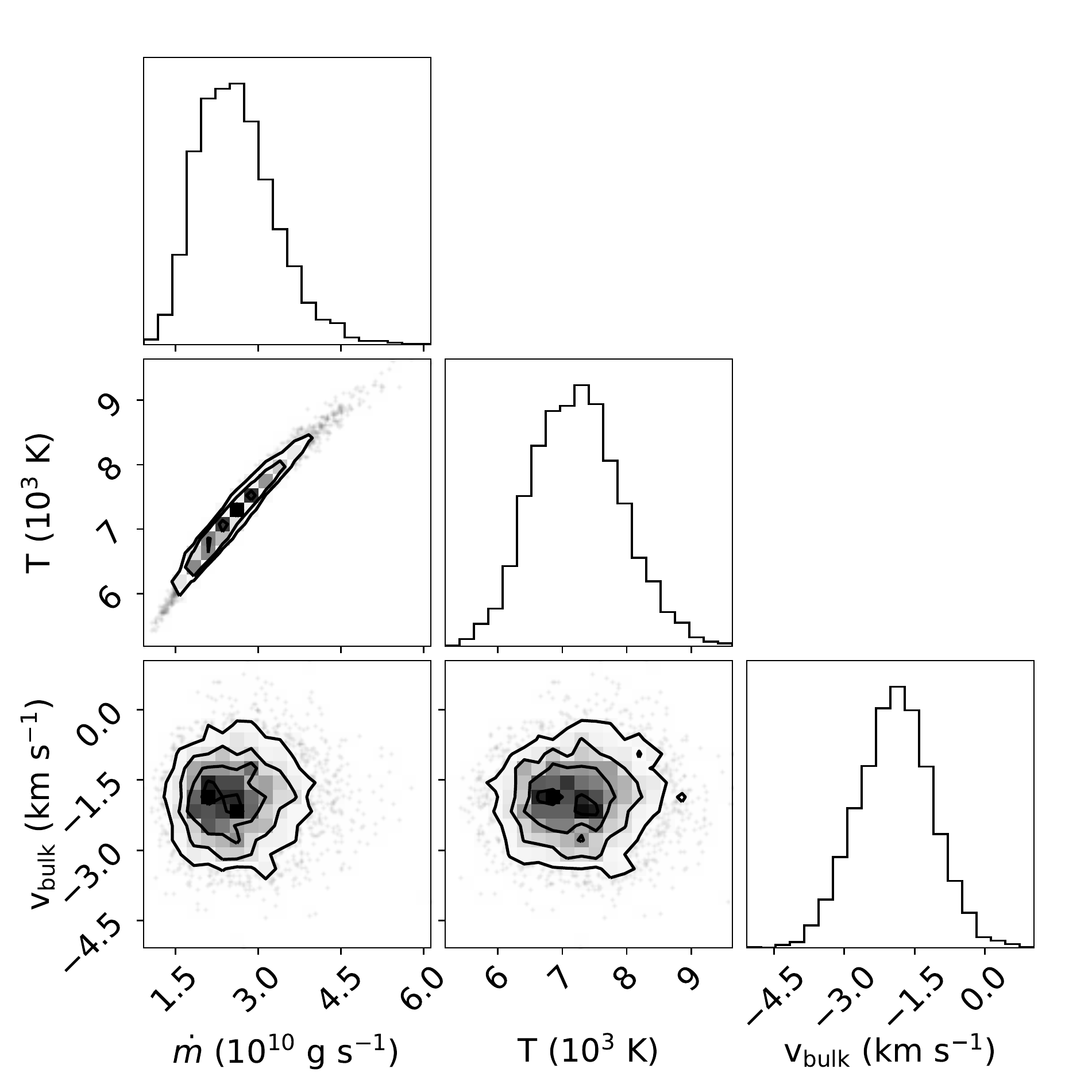}
\caption{Posterior distributions of the mass loss rate, upper atmospheric temperature, and line-of-sight bulk velocity of HAT-P-11~b using {\tt p-winds} models (no limb darkening included) as a retrieval tool against a CARMENES transmission spectrum (Model 1).}
\label{fig:hatp11b_corner}
\end{figure}

\begin{figure}
\centering
\includegraphics[width=0.9\hsize]{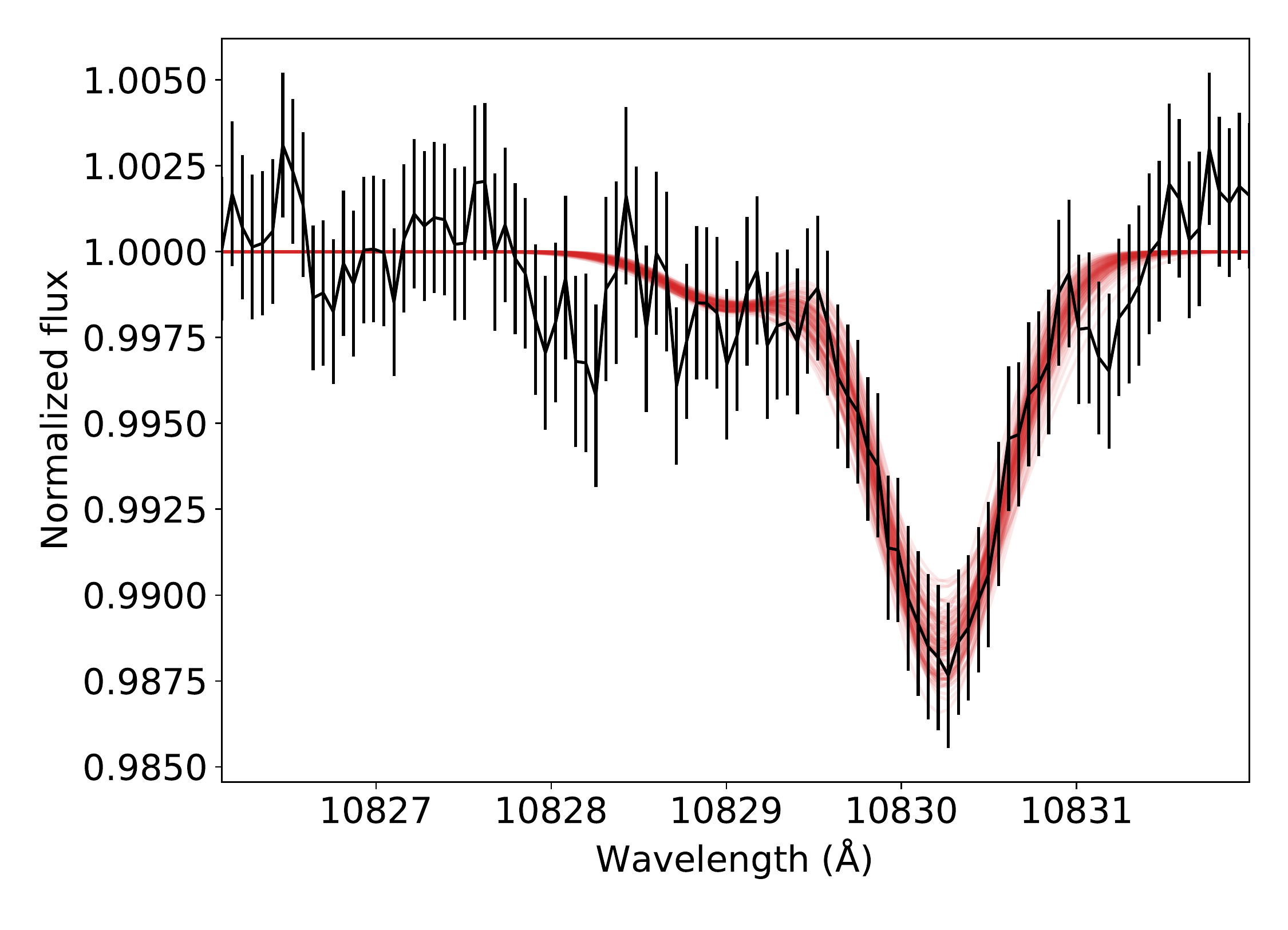}
\caption{Transmission spectrum of HAT-P-11~b measured with CARMENES (black) and a sample of 100 {\tt p-winds} models (Model 1) fit to the data (red).}
\label{fig:hatp11b_t_spec}
\end{figure}

We ran the MCMC for 7000 steps and 10 walkers in 10 cores of a computer cluster with an average frequency of 3.0 GHz per core. The autocorrelation time $t$ of the MCMC was, on average, 45 steps when we started from a first guess of $1 \times 10^{10}$~g~s$^{-1}$, $800$~K, and $-2.0$~km~s$^{-1}$, respectively, for $\dot{m}$, $T$, and $v_{\rm bulk}$. We remove a total of $2t$ burn-in steps from the beginning of the MCMC and take a sample thinned by $t / 2$, resulting in a flat chain of approximately 3600 samples. The computation of the MCMC chains took approximately 6.5 hours of computing time. Different planets will likely yield different computing times because the numerical bottleneck (calculating the He distribution) is highly dependent on the input parameters. We show the posterior distributions of the fit parameters for Model 1 in Fig. \ref{fig:hatp11b_corner} (see also Appendix \ref{app:posteriors}), and a sample of corresponding transmission spectrum models fit to the data in Fig. \ref{fig:hatp11b_t_spec}.

\begin{figure}
\centering
\includegraphics[width=0.9\hsize]{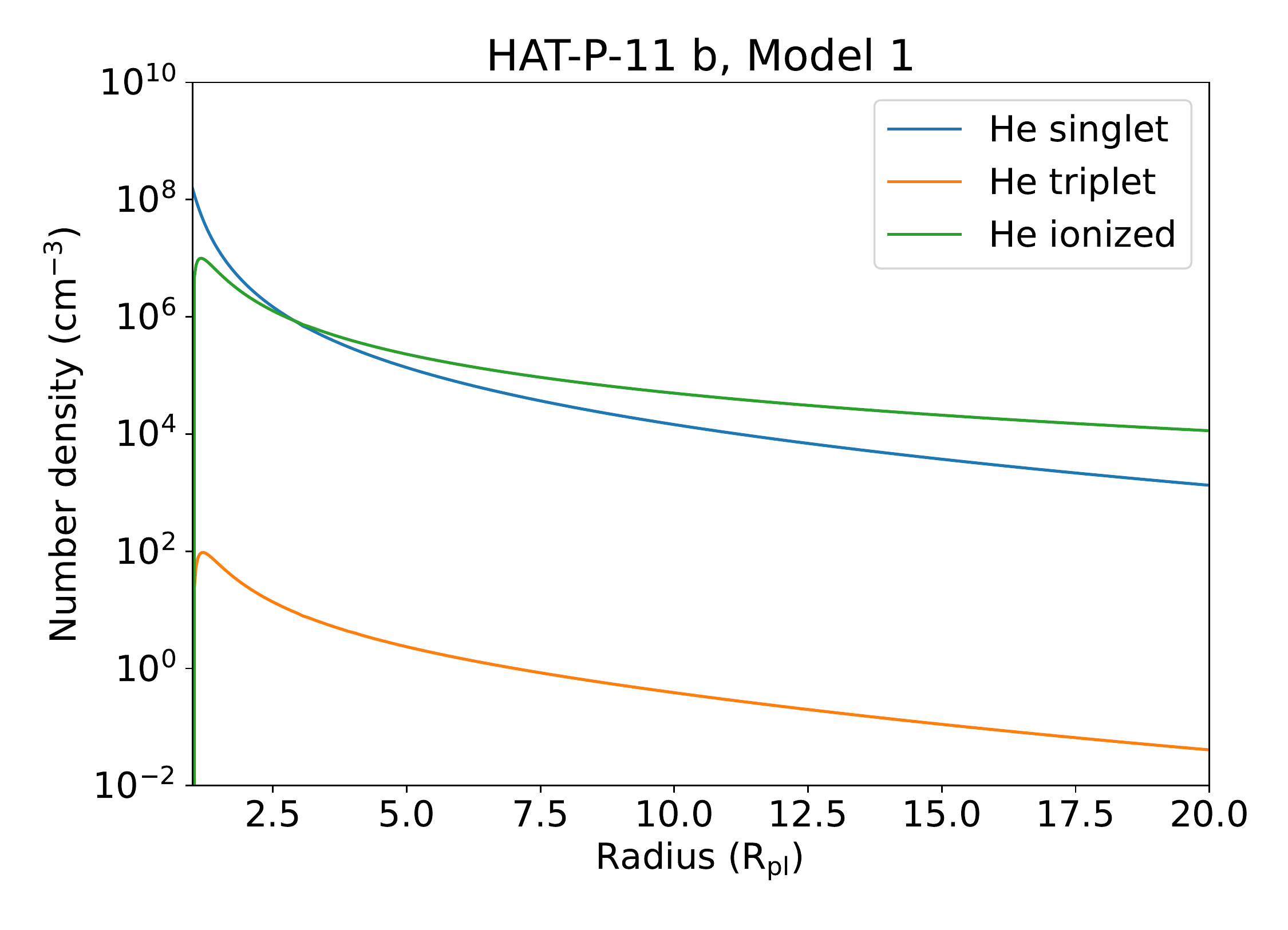}
\caption{Distribution of He in the upper atmosphere of HAT-P-11~b based on the best-fit solution obtained by fitting {\tt p-winds} models (no limb darkening included) to the CARMENES transmission spectrum (Model 1).}
\label{fig:hatp11b_He_dist}
\end{figure}

\begin{figure}
\centering
\includegraphics[width=0.9\hsize]{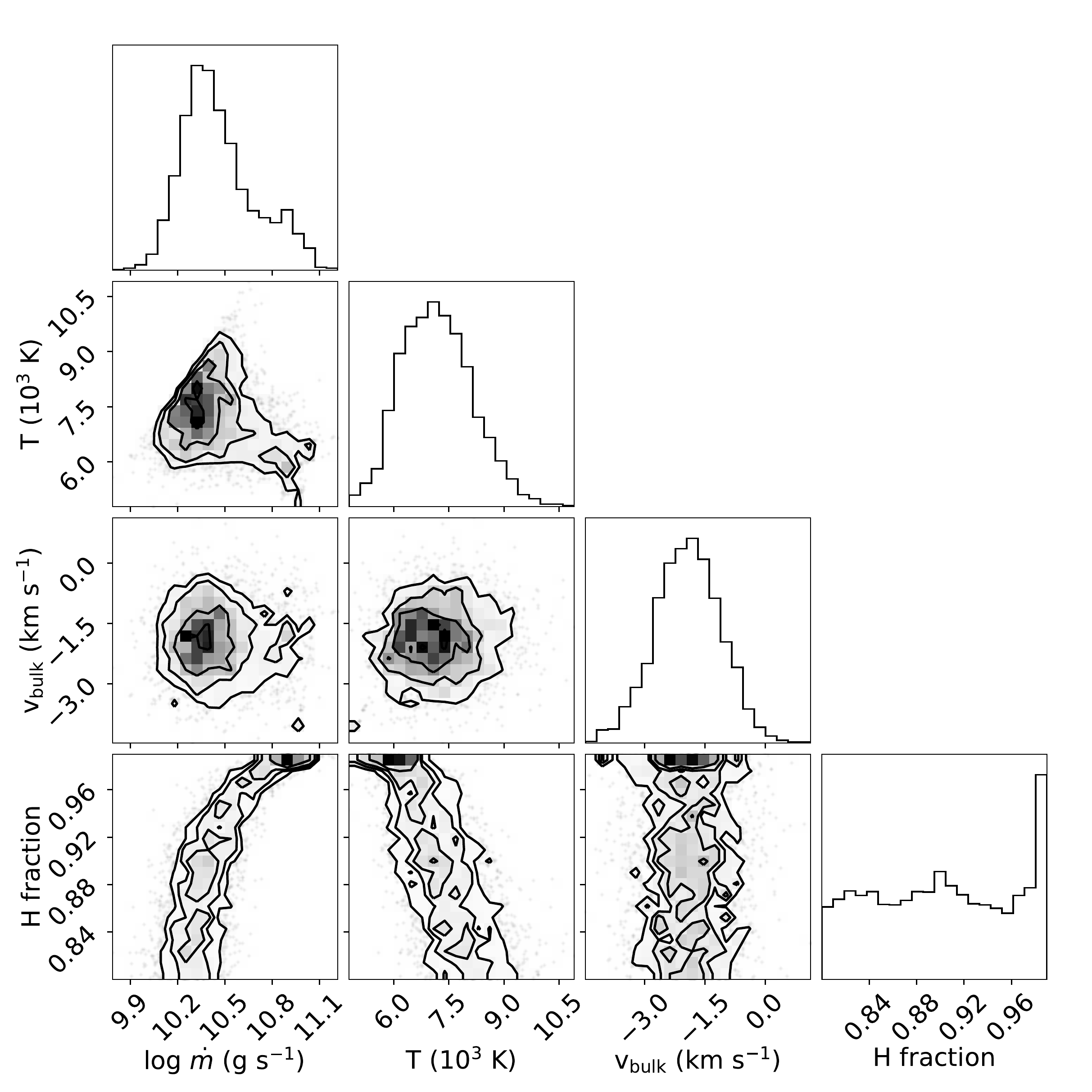}
\caption{Same as Fig. \ref{fig:hatp11b_corner}, but including the H fraction as a free parameter to be fit (Model 3).}
\label{fig:hatp11b_corner_4p}
\end{figure}

Table \ref{hp11b_results} contains the retrieved atmospheric escape parameters for HAT-P-11~b based on the CARMENES transmission spectrum. All models we tested yield results consistent with one another within their uncertainties, based on the marginalized posterior distribution of the retrieved parameters. A comparison between the results for Models 1 and 4 reveals that, in the case of HAT-P-11~b, the limb darkening of the star does not significantly affect the retrieved escape parameters when fitting a ground-based transmission spectrum. In the case of Model 3, where we allowed the H fraction to vary between 0.80 and 0.99, the retrieval slightly favors fractions $> 0.96$, but the 3$\sigma$ upper limit of $> 0.80$ is not constraining. We show the resulting posterior distributions of the fit to Model 3 in Fig. \ref{fig:hatp11b_corner_4p}. For fractions above $0.92$, the retrieved escape rate tends to increase by a factor of several percent. The retrieved upper atmosphere temperature $T$ is highly anticorrelated with the H fraction. This degeneracy increases the uncertainties of $T$ by a factor of at least two. We show the resulting distribution of He in the upper atmosphere of HAT-P-11~b in Fig. \ref{fig:hatp11b_He_dist} based on the best-fit model to the CARMENES data. Finally, we show that the retrieved escape parameters of Models 1-3 (average-velocity approximation for the wind broadening) are fully consistent with that of Model 4 (formal radiative transfer calculation). Hence, we demonstrate that this approximation, which saves an order of magnitude in computation time, does not significantly affect the retrieved escape parameters.

\begin{table*}[t]
\caption{Upper atmosphere properties retrieved for HAT-P-11~b from the CARMENES transmission spectrum.}
\label{hp11b_results}
\centering
\begin{tabular}{lcccc}
\hline\hline
Model & $\dot{m}$ & $T$ & $v_{\rm bulk}$ & H fraction \\
 & ($\times 10^{10}$~g~s$^{-1}$) & ($\times 10^{3}$~K) & (km~s$^{-1}$) & \\
\hline
1 & $2.5^{+0.8}_{-0.6}$ & $7.2 \pm 0.7$ & $-1.9 \pm 0.8$ & $0.90$ (fixed) \\
2 & $2.3^{+0.7}_{-0.5}$ & $7.2^{+0.7}_{-0.6}$ & $-1.9 \pm 0.8$ & $0.90$ (fixed) \\
3 & $2.6^{+2.6}_{-0.8}$ & $7.1^{+1.0}_{-0.9}$ & $-1.9 \pm 0.8$ & $> 0.80$ (3$\sigma$) \\
4 & $2.1$ & $6.7$ & $-1.9$ & $0.90$ (fixed) \\
\hline
\end{tabular}
\end{table*}

In order to compare our results with those obtained by the three-dimensional model EVE \citep{2013A&A...557A.124B} used by \citet{2018Sci...362.1384A}, we need to calculate the escape of metastable He only instead of total mass loss. For a total escape rate of $2.5 \times 10^{10}$~g~s$^{-1}$ and an upper atmosphere temperature of $7200$~K obtained from the retrieval described above, we calculate an average metastable helium fraction of $4.8 \times 10^{-6}$ and a $T/\mu$ fraction of 8000~K~amu$^{-1}$. This result translates into a metastable-He escape rate of $1.2 \times 10^5$~g~s$^{-1}$, which is compatible with the upper-limit rate of $\sim 3 \times 10^5$~g~s$^{-1}$ determined by \citet{2018Sci...362.1384A}. Our retrieved $T/\mu$ fraction when assuming a H fraction of 0.9 is discrepant with the results of \citet{2018Sci...362.1384A}, who \textbf{find} $T/\mu = 24\,000$~K~amu$^{-1}$. Some of the solutions of our retrieval with the H fraction as a free parameter do allow for high values of $T/\mu$ up to $12\,000$, but they are nevertheless incompatible with \citet{2018Sci...362.1384A}; the authors, however, do propose that a high $T/\mu$ may correspond to a low mean atomic weight, which can be obtained with a large fraction of ionized gas and free electrons. In an upcoming manuscript, \citet{Vissapragada2021} will discuss how solutions with high temperatures can be ruled out because they are not energetically self-consistent, assuming that the heating comes solely from the available high-energy irradiation budget. A possible explanation for the disagreement can be due to modeling difference, since \citet{2018Sci...362.1384A} use a hydrostatic model while we use a hydrodynamic one; since a hydrostatic thermosphere is less extended, it needs a higher temperature to increase the density of He enough to be detectable at high altitudes.

The bulk velocity of $-1.9 \pm 0.8$~km~s$^{-1}$ is consistent with the net blueshift of $3$~km~s$^{-1}$ reported by \citet{2018Sci...362.1384A}, which was previously interpreted as a high-altitude wind flowing from the day- to the night-side of the planet. This net blueshift is not predicted by the one-dimensional Parker wind model, which is the reason for fitting it as a free parameter in our models. More complex, tridimensional models that take into account other physical processes may be necessary to determine the exact mechanism that causes this bulk velocity shift in the metastable He absorption signature.

\subsection{Escape rate upper limit for a non-detection in GJ~436~b}

GJ~436~b is a high-profile case of atmospheric escape because it possesses the deepest transmission spectrum feature detected to date: a repeatable 50\% in-transit absorption in Lyman-$\alpha$ \citep{2014ApJ...786..132K, 2015Natur.522..459E, 2017A&A...605L...7L, 2019A&A...629A..47D}, which is explained by a large volume of exospheric neutral H fed by escape \citep{2016A&A...591A.121B, 2021MNRAS.501.4383V}. In fact, \citet{2018ApJ...855L..11O} predict a metastable He signature as deep as 9\% in the core of the strongest line of the triplet. However, when GJ~436~b was observed by CARMENES, the results yielded only a non-detection \citep{2018Sci...362.1388N}.

In this section we \textbf{attempt} to measure an upper limit of atmospheric escape rate to the non-detection of He in GJ~436~b and compare it to the result derived from Lyman-$\alpha$ transmission spectroscopy and modeling. The metastable He transmission spectrum of GJ~436~b is rather unfortunately not publicly available, but the pipeline-reduced spectral time series from \citet{2018Sci...362.1388N} is available in the CARMENES data archive\footnote{\footnotesize{\url{http://carmenes.cab.inta-csic.es/gto/jsp/nortmannetal2018.jsp}}}.

We ran an MCMC of $10\,000$ steps and 10 walkers with three free parameters: total escape rate, upper atmospheric temperature, and the H fraction. We increased the number of steps compared to the HAT-P-11~b retrieval in order to better explore the parameter space, since we are expecting to obtain only upper/lower limits. We did not include limb darkening. Based on previous theoretical predictions for GJ~436~b \citep[e.g.,][]{2016A&A...586A..75S}, we set uniform priors of [$10^7$, $10^{12}$]~g~s$^{-1}$ for the mass loss and [$1\,000$, $10\,000$]~K for the temperature, and $[0.40, 0.99]$ for the H fraction. We used the high-energy spectrum of GJ~436 measured in the MUSCLES Treasury Survey as a source of irradiation.

The resulting posterior distributions of the free parameters for GJ~436~b yield, at 99.7\% (3$\sigma$) confidence, an upper limit of $4.5 \times 10^9$~g~s$^{-1}$ for the escape rate and a lower limit of 2600~K for the upper atmospheric temperature (given the uniform priors above). In broad terms, mass loss rates above this value or temperatures below the lower limit would yield a detectable metastable He signature. With the H fraction as a free parameter, these 3-$\sigma$ limits become $3.4 \times 10^{10}$~g~s$^{-1}$ and 2400~K. This result is consistent with the escape rate of $\sim 2.5 \times 10^8$~g~s$^{-1}$ inferred by \citet{2016A&A...591A.121B}, and with the mass loss rate of $(6 - 10) \times 10^9$~g~s$^{-1}$ inferred by \citet{2021MNRAS.501.4383V}, both based on the same Lyman-$\alpha$ transmission spectroscopy data set.

Given the flat prior of $[0.40, 0.99]$ for the H fraction, the resulting posterior distribution of this parameter is not constraining; however, it seems to favor higher values and peaks near 0.99 (see Fig. \ref{fig:gj436b_post2}). Interestingly, this result could be seen as an agreement with the prediction of a H-rich outflow for GJ~436~b due to selective escape, leading to a He-rich lower atmosphere \citep{2015ApJ...807....8H}. We cannot, however, draw strong conclusions on this matter because GJ~436~b had a non-detection of He. More detailed descriptions that fit both H and He simultaneously in the upper atmosphere of this planet are likely going to yield more definitive answers. For example, \citet{2021A&A...647A.129L} used H densities derived from Lyman-$\alpha$ observations to inform metastable He models, and determine that the warm Neptune GJ~3470~b has $n_{\rm H}/n_{\rm atoms} = 0.985^{+0.010}_{-0.015}$.

\section{Conclusions}\label{sect:conclusions}

We demonstrate in this manuscript the usage of the open-source Python code {\tt p-winds} to forward model the distribution of He atoms in the upper atmospheres of exoplanets as well as their corresponding metastable He transmission spectra of exoplanets. The code also enables the retrieval of atmospheric escape rates and temperatures based on observations at high resolution when coupled to an optimization algorithm, such as a maximum likelihood estimation or an MCMC sampler. A typical retrieval takes several hours to compute, depending on the setup. 

As an implementation of the method originally described by \citet{2018ApJ...855L..11O}, the forward models produced by {\tt p-winds} are fully compatible with that study. We also implement changes proposed by \citet{2020A&A...636A..13L}, such as the inclusion of charge exchange of He and H particles. Our implementation includes further improvements, such as the addition of transit geometry and the limb darkening of the host star, as well as allowing the H fraction ($n_{\rm H} / n_{\rm atoms}$) as a free parameter in the retrieval.

We used {\tt p-winds} to fit the escape rate, outflow temperature, and the H fraction of the warm Neptune HAT-P-11~b based on CARMENES transmission spectroscopy previously reported in \citet{2018Sci...362.1384A}. For a model without limb darkening and with $n_{\rm H}/n_{\rm atoms}$ fixed at 0.90, we find that the escape rate of HAT-P-11~b is $(2.5^{+0.8}_{-0.6}) \times 10^{10}$~g~s$^{-1}$ and the planetary outflow temperature is $7200 \pm 700$~K. This temperature is in disagreement with the value of $T/\mu$ calculated by \citet{2018Sci...362.1384A}, and it is likely caused by a key difference between our models -- theirs contains a hydrostatic thermosphere, while ours is hydrodynamic. Including limb darkening does not have a significant impact on the retrieved parameters of HAT-P-11~b. Allowing the H fraction as a free parameter has a stronger impact because it yields an anticorrelation with the retrieved outflow temperature. It also increases the uncertainty of the retrieved atmospheric escape rate. We find that the H fraction is unconstrained, but with a preference for higher values. These results are in agreement with those of \citet{2020A&A...636A..13L, 2021A&A...647A.129L}, although those authors can constrain the H fraction by analyzing He transmission spectra in conjunction with H escape using Lyman-$\alpha$ observations.

Finally, we also attempted to fit limits for the escape rate, outflow temperature, and H fraction of GJ~436~b based on a non-detection with the CARMENES spectrograph reported by \citet{2018Sci...362.1388N}. We find an upper limit of $3.4 \times 10^{10}$~g~s$^{-1}$ for the first and a lower limit of 2400~K for the second at 99.7\% confidence. Our upper and lower limit determinations show a preference for high values of $n_{\rm H}/n_{\rm atoms}$, with the posterior distribution peaking near 0.99. These results are fully compatible with the escape rate of $\sim 2.5 \times 10^8$~g~s$^{-1}$ inferred by \citet{2016A&A...591A.121B} based on Lyman-$\alpha$ transmission spectroscopy. For both HAT-P-11~b and GJ~436~b, we find a slight preference for high values of H in the atomic fraction, which is in line with the results of \citet{2020A&A...636A..13L, 2021A&A...647A.129L} for other hot gas giants.

The main limitations of a one-dimensional, isothermal Parker wind model are: 1) It does not capture the three-dimensional nature of very extended atmospheres, particularly when they have both a thermospheric and an exospheric contributions \citep[see the case of WASP-107~b in][]{2019A&A...623A..58A, 2021arXiv210708999S}; 2) it does not take into account the variable profile of temperature with radial distance from the planet, which is seen in self-consistent models of escape \citep[e.g.,][]{2016A&A...586A..75S, 2019MNRAS.490.3760A}; and 3) it does not self-consistently consider the sources of heating and cooling that control the atmospheric escape process. The usefulness of simple models such as {\tt p-winds} lies in an efficient exploration of the parameter space that defines atmospheric escape (scalability) and ease of use (open-source, fully documented code) when more sophisticated models are not yet warranted.

As for the next steps, we aim to improve {\tt p-winds} by including the escape of heavier atomic species, such as C, N, O, Mg, Si, and Fe. This will allow us to use the code to predict and interpret observations of metals escaping hot gas giants, such as the signatures reported by \citet{2013A&A...560A..54V} and \citet{2019AJ....158...91S}. We shall also add day-to-nightside winds to the atmospheric modeling, similar to \citet{2020A&A...633A..86S}. Another avenue to explore {\tt p-winds} in the future consists in coupling it with more complex tridimensional hydrodynamic escape models.

\begin{acknowledgements}
LADS acknowledges the helpful input of A. Wyttenbach, M. Stalport, A. Oklop{\v{c}}i{\'c}, J. St\"urmer, and M. Zechmeister to the development of this project. The authors also thank the referee, Manuel L\'opez-Puertas, for the helpful and detailed review. SV is supported by an NSF Graduate Research Fellowship and the Paul \& Daisy Soros Fellowship for New Americans. RA is a Trottier Postdoctoral Fellow and acknowledges support from the Trottier Family Foundation, and his contribution was supported in part through a grant from {\it Fonds de recherche du Québec – Nature et technologies}. This research was enabled by the financial support from the European Research Council (ERC) under the European Union's Horizon 2020 research and innovation programme (projects: {\sc Four Aces} grant agreement No 724427; {\sc Spice Dune} grant agreement No 947634; {\sc ASTROFLOW} grant agreement No 817540), and it has been carried out in the frame of the National Centre for Competence in Research PlanetS supported by the Swiss National Science Foundation (SNSF). The {\tt p-winds} code makes use of the open source software NumPy \citep{harris2020array}, SciPy \citep{2020SciPy-NMeth}, Pillow (\url{https://python-pillow.org}), and Astropy \citep{astropy:2018}. The results of this manuscript were also made possible by the open source software Matplotlib \citep{Hunter:2007}, OpenMPI (\url{https://www.open-mpi.org}), Jupyter \citep{jupyter}, MPI for Python \citep[{\tt mpi4py};][]{DALCIN20111124}, {\tt emcee} \citep{2013PASP..125..306F}, and {\tt schwimmbad} \citep{schwimmbad}. Finally, the authors also extend a special thanks to the platforms GitHub, Conda-Forge, Read the Docs, and Travis.ci for the valuable support of open-source initiatives.
\end{acknowledgements}

\bibliographystyle{aa}
\bibliography{references.bib}

\appendix

\section{The {\tt flatstar} code}\label{app:flatstar}

The code implemented in {\tt flatstar} was originally written as a part of the {\tt transit} module of {\tt p-winds}. However, we decided to transform it into a separate package because this implementation can be useful for other astrophysical applications not necessarily related to transmission spectroscopy. 

The typical usage of {\tt flatstar} involves setting a grid shape $(N_{\rm x}, N_{\rm y})$ and stellar radius $R_{\rm s}$ in number of pixels, and optionally setting a limb-darkening law. The limb-darkening laws currently implemented in the code are: linear, quadratic \citep{1950HarCi.454....1K}, square-root \citep{1992A&A...259..227D}, logarithmic \citep{1970AJ.....75..175K}, exponential \citep{2003A&A...412..241C}, the three-parameter law of \citet{2009A&A...505..891S}, and the four-parameter law of \citet{2000A&A...363.1081C}. Finally, the user can also set a custom limb-darkening law. The star is always centered to the grid.

In addition, {\tt flatstar} can add a planetary transit with user-defined planet-to-star ratio, $R_{\rm p}/R_{\rm s}$, transit impact parameter, $b$, and phase, $\phi$. The first and fourth contact of the transit are defined as the phases $-0.5$ and $+0.5$, respectively, independent of $b$. The y coordinate of the planetary center ($y_{\rm p}$) in pixel space is calculated as

\begin{equation}
    y_{\rm p} = (b \times R_{\rm s}) + N_{\rm y} / 2 \mathrm{.}
\end{equation}The x coordinate of the planetary center ($x_{\rm p}$) is not as straightforward to calculate, since we define it based on $\phi$ and $b$. Let $\theta$ be the angle between the $y$ axis and the vector that connects the center of the star and the planet at first contact:

\begin{equation}
    \cos{\theta} = \frac{b \times R_{\rm s}}{R_{\rm p} + R_{\rm s}}\mathrm{.}
\end{equation}The distance $\beta$ from the planet center to the $y$ axis at first contact is given by

\begin{equation}
    \beta = \left( R_{\rm p} + R_{\rm s} \right) \sin{\theta} = \left( R_{\rm p} + R_{\rm s} \right) \sqrt{1 - \left(b \times \frac{R_{\rm s}}{R_{\rm p} + R_{\rm s}} \right)^2} \mathrm{.}
\end{equation}Thus, the distance $x_0$ of the planet center from the border of the simulation at first contact is

\begin{equation}
    x_0 = N_{\rm x} / 2 - \beta \mathrm{.}
\end{equation}As the planet moves from the first contact to fourth contact, it covers a distance of $2\beta$. For arbitrary phases between $\phi = -0.5$ (first contact) and $\phi = +0.5$ (fourth contact), the distance $x_p$ of the planet from the border of the simulation is thus

\begin{equation}
    x_{\rm p} = x_0 + 2\beta \times (\phi + 0.5) \mathrm{.}
\end{equation}This formulation assumes that the arc that the planet follows during the transit can be approximated to a chord.

The grid can be super-sampled in order to avoid "hard" pixel edges when the grid size is coarse. This is useful to save computation time in cases where we need to mass produce grids while conserving the precision of intensities (which is the case of atmospheric retrievals with {\tt p-winds}). By default, the resampling algorithm is the "box" method, which takes the value of each pixel with fixed weights to compute the average flux of the resampled pixel. We do not recommend using {\tt flatstar} to fit wide band photometric light curves, since the computation time is much longer than other codes that implement analytical equations to calculate light curves, such as {\tt batman} \citep{2015PASP..127.1161K}.

\section{Other posterior distributions for the fits to HAT-P-11~b and GJ~436~b data}\label{app:posteriors}

\begin{figure}[h]
\centering
\includegraphics[width=0.9\hsize]{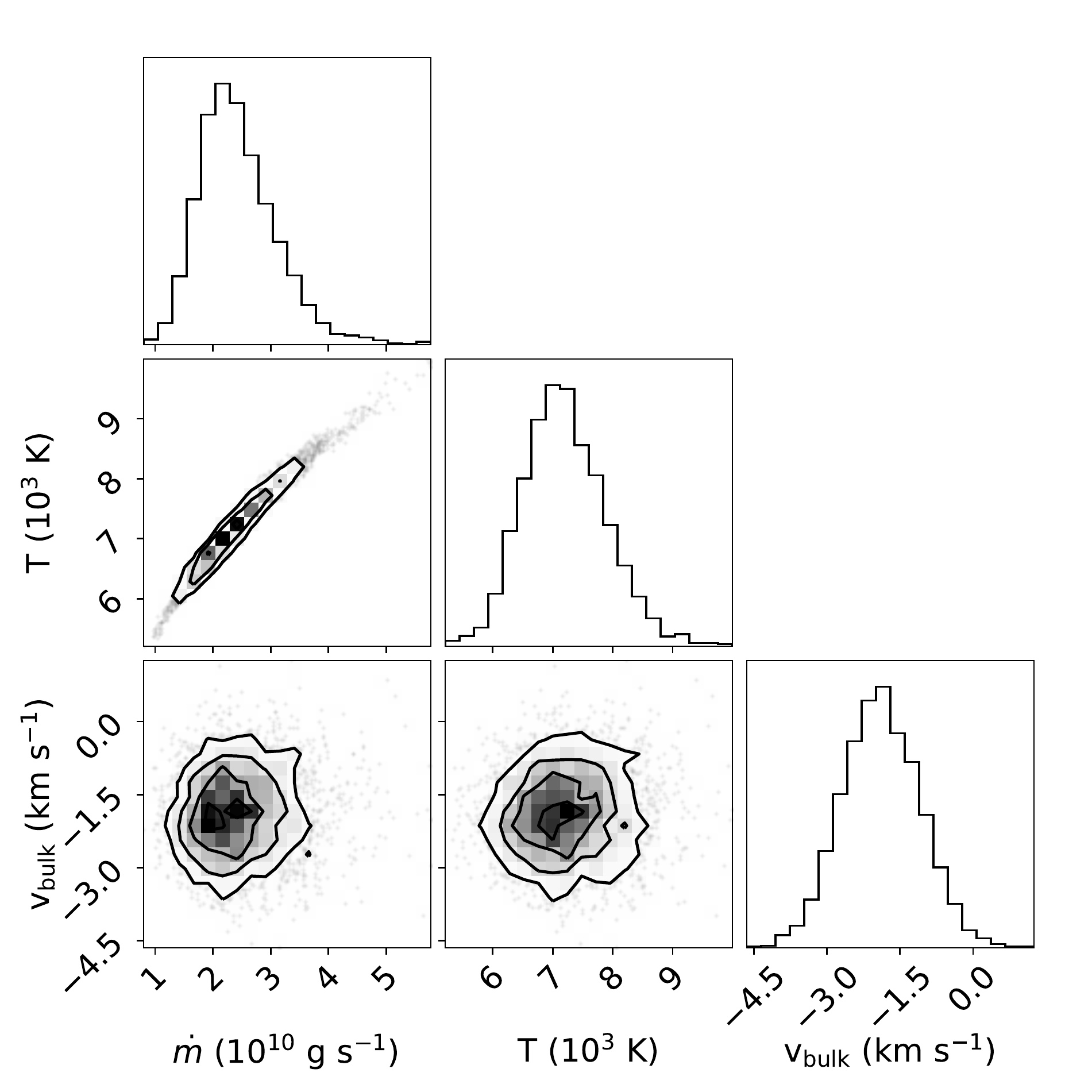}
\caption{Posterior distribution of parameters of HAT-P-11~b fit to the CARMENES transmission spectrum using a model with a quadratic limb-darkening law and coefficients $c_1 = 0.63$ and $c_2 = 0.09$.}
\label{fig:hp11b_post2}
\end{figure}

\begin{figure}[h]
\centering
\includegraphics[width=0.7\hsize]{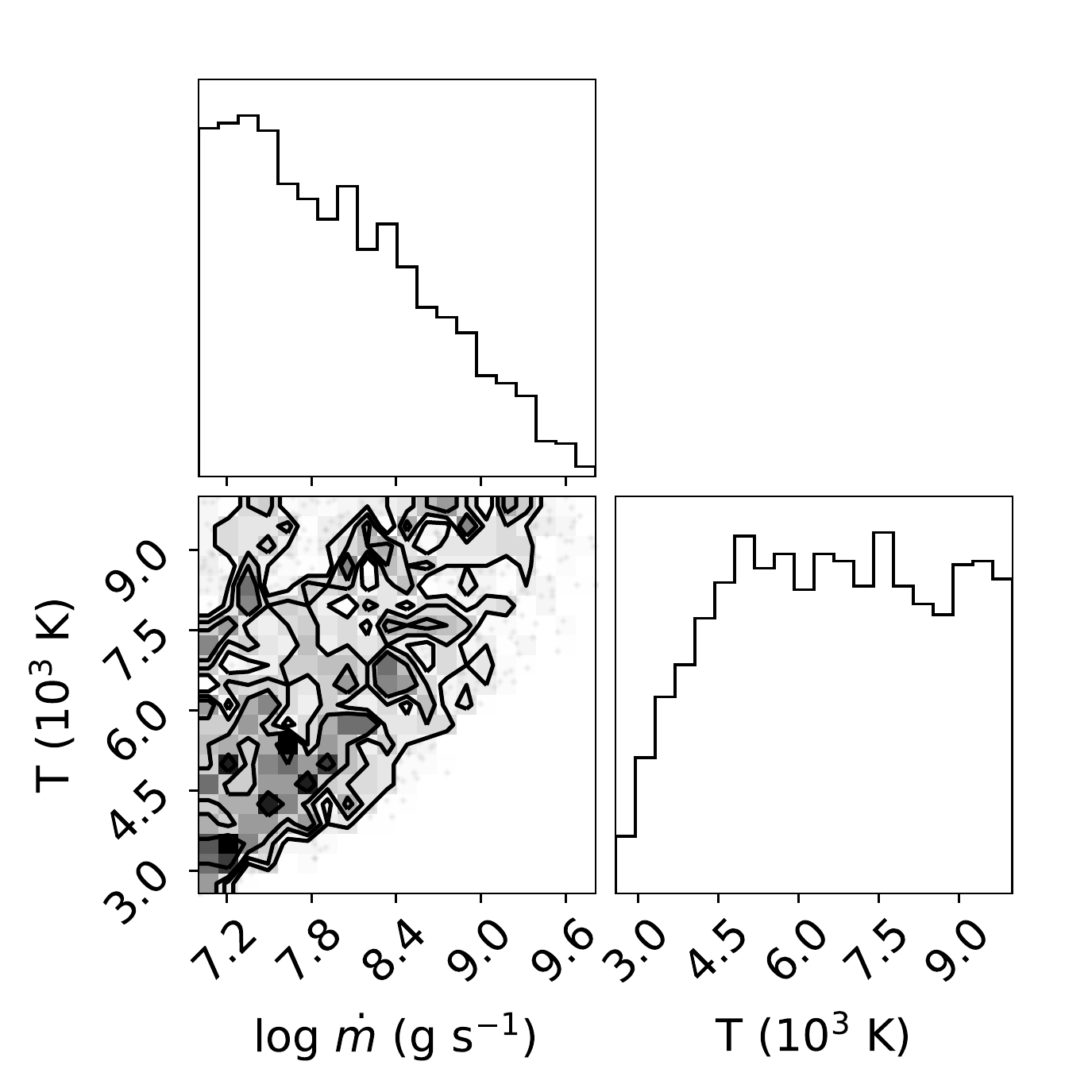}
\caption{Same as Fig. \ref{fig:hp11b_post2}, but for GJ~436~b with the H fraction fixed to 0.90 and no limb darkening.}
\label{fig:gj436b_post}
\end{figure}

\begin{figure}[h]
\centering
\includegraphics[width=0.9\hsize]{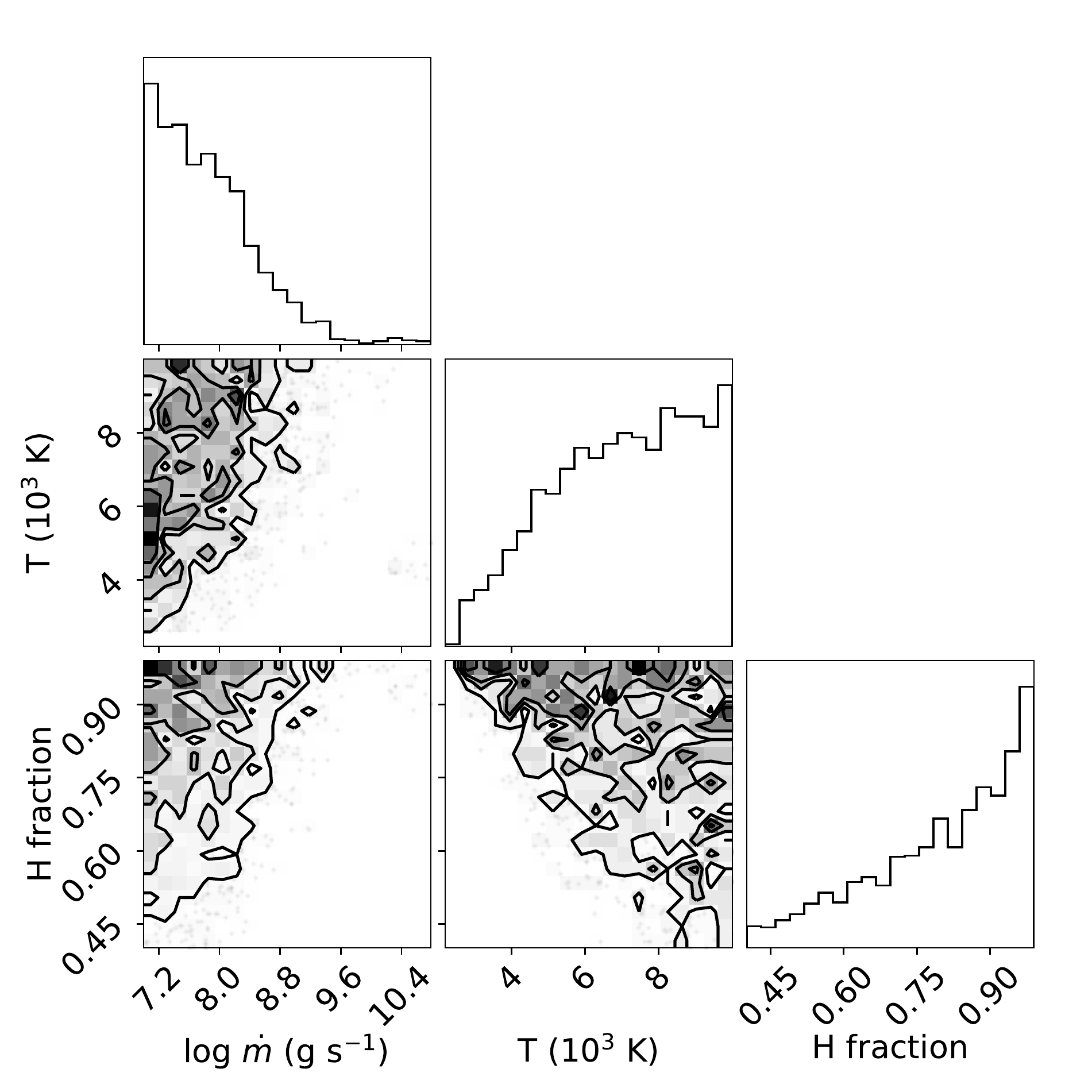}
\caption{Same as Fig. \ref{fig:gj436b_post}, but with the H fraction as a free parameter.}
\label{fig:gj436b_post2}
\end{figure}

\end{document}